\documentclass[10pt,journal,compsoc]{IEEEtran}
\ifCLASSOPTIONcompsoc
  \usepackage[nocompress]{cite}
\else
  \usepackage{cite}
\fi
\usepackage{amsmath}
\usepackage{hyperref}
\usepackage{graphicx}
\usepackage{epsfig}
\usepackage{subfig}

\ifCLASSINFOpdf
\else
\fi
\begin{document}
%
\title{\huge{Evaluation of Interpretability for Deep Learning algorithms  in EEG Emotion Recognition: A case study in Autism}}
%
%
%
%

\author{Juan Manuel Mayor-Torres, Sara Medina-DeVilliers, Tessa Clarkson, Matthew D. Lerner, and Giuseppe Riccardi ~\IEEEmembership{Member and Fellow,~IEEE}
\IEEEcompsocitemizethanks{\IEEEcompsocthanksitem Juan Manuel Mayor-Torres** and Giuseppe Riccardi are with the Department of Information Engineering and Computer Science, University of Trento, Via Sommarive, Povo, Trento, 1328, Italy. E-mail: juan.mayortorres@unitn.it, giuseppe.riccardi@unitn.it\\
\IEEEcompsocthanksitem Sara Medina-Devilliers and Matthew D. Lerner are with the Department of Psychology, StonyBrook University NY, USA, E-mail sara.medina-devilliers@stonybrook.edu, matthew.lerner@stonybrook.edu \\
\IEEEcompsocthanksitem Tessa Clarkson is with the Department of Psychology, Temple University, Philadelphia, PA email: tessa.clarkson@temple.edu
}
\thanks{}}

%
%

\markboth{}%
{Mayor-Torres \MakeLowercase{\textit{et al.}}: Bare Demo of IEEEtran.cls for Computer Society Journals}
%



\IEEEtitleabstractindextext{%
\begin{abstract}
 Current models on Explainable Artificial Intelligence (XAI) have shown an evident and quantified lack of reliability for measuring feature-relevance when statistically entangled features are proposed for training deep classifiers. There has been an increase in the application of Deep Learning in clinical trials to predict early diagnosis of neuro-developmental disorders, such as Autism Spectrum Disorder (ASD). However, the inclusion of more reliable saliency-maps to obtain more trustworthy and interpretable metrics using neural activity features is still insufficiently mature for practical applications in diagnostics or clinical trials. Moreover, in ASD research the inclusion of deep classifiers that use neural measures to predict viewed facial emotions  is relatively unexplored. Therefore, in this study we propose the evaluation of a Convolutional Neural Network (CNN) for electroencephalography (EEG)-based facial emotion recognition decoding complemented with a novel RemOve-And-Retrain (ROAR) methodology to recover highly relevant features used in the classifier. Specifically, we compare well-known relevance maps such as Layer-Wise Relevance Propagation (LRP), PatternNet, Pattern-Attribution, and Smooth-Grad Squared. This study is the first to consolidate a more transparent feature-relevance calculation for a successful EEG-based facial emotion recognition using a within-subject-trained CNN in typically-developed and ASD individuals. 
\end{abstract}

}

\maketitle

\IEEEdisplaynontitleabstractindextext

%
\IEEEpeerreviewmaketitle

\IEEEraisesectionheading{\section{Introduction}\label{sec:introduction}}

%
%
%
%
\IEEEPARstart{D}{eep} Learning (DL) has led to large improvements in many domains: e.g. image recognition \cite{krizhevsky2012imagenet}, automated translation \cite{sutskever2014sequence}, and object detection \cite{he2017mask}. In recent years, researchers have started to investigate DL for clinical application. For example: the use of DL for the diagnosis of complex neurodevelopmental disorders, such as,  Parkinson \cite{prince2019evaluation}, Rett \cite{mustafamultimodal}, or Alzheimer \cite{andreotti2018multichannel}.  Other uses include facial emotion recognition (FER) in typically-developed (TD) individuals \cite{liu2016emotion, li2018cross,weitz2018towards,ghoshal2019estimating} and individuals with ASD \cite{torres2021facial}.
Despite this, most research on processing electroencephalography (EEG) data still relies on more traditional machine approaches such as Support Vector Machine (SVM) and Linear Discriminant Analysis (LDA). This includes the work on ASD diagnoses \cite{bosl2011eeg,castelhano2018stimulus} and FER \cite{jenke2014feature,koelstra2011deap, fan2017eeg, fan2017eeg2}. These methods were quite successful in multiple applications such as motor imagery decoding \cite{blankertz2011single} and artifact removal \cite{winkler2011automatic}. However, their accuracy is limited for more complex applications such as FER decoding.

Traditional classifiers necessitate explicit feature extraction prior to classification in order to achieve high accuracy \cite{o2017classification,torres2013eeg,mayor2018t}. In the aforementioned case of motor imagery decoding, features were generated using Common-Spatial Patterns (CSP) which is directly linked to neurophysiological processes \cite{yang2015use}. Unfortunately, in the case of FER decoding relevant features are not known or directly linked to neurophysiological data \cite{ang2008filter,lee2014classifying,blankertz2004bci}. This limitation motivates use of DL, such as CNNs, to process minimally preprocessed EEG data and recover features \cite{torresenhanced,schirrmeister2017deep}.

To our knowledge, our previous study is the first successful application of DL classifiers on neural activity to decode facial  emotions in ASD populations  \cite{torres2021facial}.  In this study, we constructed a 2D image from the EEG as an input by stacking the individual EEG channels vertically - columns represent time and rows represent different EEG channels so that a single EEG image  can be processed directly by a CNN \cite{torres2021facial,schirrmeister2017deep}. One could think of these EEG images as scribble drawings made by a child who is learning to draw. 

Each scribble drawing represents a different encoded facial emotion in the neural activity or a single-trial. In other words, the CNN would need to find the best class separability using the features that are specific to each scribble drawing or facial emotion. What makes this task particularly difficult is that the representations are highly non-deterministic and noisy, just as the scribble drawings from a child. Consequently, understanding which features were used to classify facial emotions for typically developed (TD) controls or ASD participants performing FER \cite{torres2021facial,black2017mechanisms} is difficult.  This contributes to the CNN being an inaccessible black-box system, which can not be explained directly from the neural features or the DL models we commonly use \cite{kovalerchuk2018toward}. 

To alleviate this problem, Explainable AI (XAI) methods (i.e, saliency-maps) have been introduced recently as a way to understand DL classifiers. Some well-know XAI methods, such as  Grad-CAM \cite{selvaraju2017grad}, Grad-CAM++ \cite{chattopadhay2018grad}, Integrated Gradients (IG) \cite{kim2019saliency}, and Smooth-Grad \cite{smilkov2017smoothgrad},  were developed for image classification and semantic segmentation \cite{he2017mask,lee2014classifying}. It is not clear how applicable these methods are to decoding emotions from EEG images  \cite{kindermans2017reliability,samek2020toward}. A key difference between image classification and EEG image decoding is, as mentioned before, the presence of noise. For these reasons it is not clear which XAI methods are best suited for recovering features of the decoded EEG image. 
  
In this study, we utilize a CNN  to classify emotions FER using  neural activity in the form of an EEG image.   Our previous work demonstrated that a CNN was able to decode neural activity during an FER task in TD and ASD individuals, suggesting intact FER encoding. The discrepancy between encoded FER and behavioral performance on the FER task in ASD suggests that impairments arise as a result of problems deploying properly encoding facial emotion information into behavioral responses \cite{torres2021facial,black2017mechanisms}.  To better understand when and how FER is encoded from the neural activity in TD and ASD, we can use XAI methods to recover relevant features.

To recover what EEG features are necessary  to obtain an accurate facial emotion classification, we must first understand which XAI methods are reliable. With this goal, we analyze the following methods: Layer-Wise Relevance Propagation (LRP) \cite{binder2016layer,montavon2018methods}, PatternNet, Pattern-Attribution \cite{Kindermans2017}, and Smooth-Grad Squared \cite{adebayo2018sanity} using an approach called RemOve-And Retrain (ROAR) \cite{hooker2018evaluating}. ROAR works by systematically removing features, indicated to be informative according to the XAI methods, one a time from the CNN and obtaining classifier accuracies without that features. If after their feature-removal the classifier cannot obtain a high categorization accuracy, then the feature identified by XAI methods is indeed informative, reliable and necessary for  decoding. This approach allows us to recover which XAI methods are \emph{definitely reliable} for classifying  correct facial emotion using features from neural activity.  
  
This study is the first to evaluate reliable XAI methods using ROAR, including EEG data from TD and ASD groups, and comparing the final CNN-FER metrics with the metrics associated with the FER (behavioral performance) task.

This paper will be structured as follows: (1) In the first part of the methods section, we will describe the demographics of the TD and ASD samples, the EEG pre-processing methods used for artifact-removal and signal processing, and the CNN architecture and training. (2) In the second part of the methods section, we will describe ROAR and discuss the XAI methods we use in the ROAR evaluation. (3) In the results section, we will report the comparisons between the different XAI methods, and (4) finally, we will discuss  our conclusions in the context of ASD, and Machine Learning (ML) research. This study will provide important contributions for evaluating XAI methods using ROAR, and its application for decoding FER in individuals with and without ASD \cite{dawson2005understanding,dawson2007development}.
\vspace{-0.2cm} 
\section{Materials and Methods}
In the first subsection, we describe the participant samples and the corresponding demographics.  In the second subsection, we describe that the FER task is completed while undergoing  EEG. In the third subsection, we describe the EEG data pre-processing, artifact removal, and whitening using Zero-Component-Analysis (ZCA)  procedures implemented prior to training the CNN. In the fourth subsection, we outline the ROAR methodology and the XAI methods for feature evaluation via  feature-relevance calculation.
\vspace{-0.2cm}
\subsection{Participants}
Eighty-eight participants (Age: 15.34$\pm$1.58 years), were included in the following analyses and taken from a larger study on emotional and social processing \cite{torres2021facial,nowicki2000manual}. Of these, 48 participants (29 male; Age: 15.39$\pm$1.55 years) were TD, and 40 participants (32 male; Age: 14.77$\pm$2.16 years) had a diagnosis of ASD. All participants with an ASD diagnosis were confirmed using the Autism Diagnostic Observation Schedule, Version 2.0 (ADOS-2; ASD severity Comparison Score ADOS-CS: TD 3.33$\pm$2.71, ASD 8.15$\pm$2.05) \cite{lord2000autism} and were considered high-functioning on the Kaufman Brief Intelligence Test-2 (KBIT-2). ASD participants had significantly elevated ADOS-CS compared to the TD group (p$=$0.038), but there was no significant difference in intellectual functioning (p$=$0.227). 
\vspace{-0.2cm}
\subsection{Face Emotion Recognition (FER)}
Participants completed a web-based FER task while undergoing EEG \cite{black2017mechanisms,dawson2002neural}. During this task; participants viewed emotional facial expressions from 48 children and adults’ face photographs. These facial images were taken from the DANVA-2 image-set \cite{nowicki2000manual}. We presented the emotional faces randomly with a cross-fixation of 200 ms, and a 2-second length face presentation followed by an emotion labeling menu. The emotion labeling menu presented the face-photograph again with four basic emotion labels on the bottom (i.e., happy, sad, anger, and fear). 
   
From the DANVA-2 image-set, 24 faces showed adults portraying a particular emotion, and the other 24 showed children portraying the same emotion, resulting in six stimuli for each emotion and for each modality (i.e, children/adult).  Each EEG trial is associated with a single facial emotion stimulus presented per subject. For this study, we consider the performances obtained from this task as a human behavioral performance quota - see section 2.6 for a more detailed description of these metrics. Each participant's behavioral performance was calculated after all 48 faces were presented. The behavioral performance metrics include error-rates, accuracies, and reaction-times per participant. 

EEG trials were sorted by  facial emotion, combined and stored as a single   EEGlab \cite{delorme2004eeglab} structure for each  emotion, resulting in 4 EEGlab structures (happy, sad, anger, and fear) per participant, comprising 12 EEG trials of both adults and children stimuli.
\vspace{-0.2cm}
\subsection{EEG recordings}
EEG neural activity was recorded using a Brain Products 32-channels Brain-Vision ActiCHAmp device with an original sample rate at 1KHz. Each 2D EEG image was then composed of 752 points on the time-domain and 30 channels in the spatial domain. EEG data were digitized at 16-bit resolution. The raw EEG signal was filtered with a notch filter at 60Hz with a half-power cut-off of 12db/Oct. Each active electrode was measured online with respect to a common mode-sense active electrode producing a monopolar (non-differential) channel. The EEG data collection procedures of this study adhered to best practices for EEG data collection in ASD \cite{webb2015guidelines}.

EEG trials were segmented between -200-1500ms and downsampled to 500Hz to avoid classifier overfitting \cite{blankertz2004bci,blankertz2011single}.  Each EEG trial was re-referenced  to the T9-T10 bilateral reference \cite{webb2015guidelines}. After the re-reference process, we used the remaining 30 channels (i.e., FT9, F7, FC5, FP1, FZ, FP2, F4, F8, FC6, FT10, F4, F3, FC1, C3, FC1, FC2, C4, T7, CP5, CP1, Cz, CP2, P4, P8, CP6, T8, P7, P3, Pz, O1, O2, and Oz) to create an EEG image composed of the 30 channels $\times$ 752 time-points for each trial.
\vspace{-0.2cm}
\subsection{EEG pre-processing}
Each EEG trial was automatically cleaned using the following processes in sequence: 1) the Koethe’s \emph{cleanraw} Artifact Subspace Reconstruction (ASR) in Prep pipeline complemented by the Makoto’s pipeline \cite{bigdely2018finding} for bad channels removal, and 2) the ADJUST EEGlab plugin for blinking and movement artifact removal using the 2 electro-occulogram (EOG) channels \cite{mognon2011adjust}.
  
These processes were applied with the purpose of excluding corrupted or distorted channels with artifacts, signal dropout, and electrode malfunction before using ADJUST. The ADJUST plugin uses spatial and temporal features such as temporal kurtosis, spatial average-difference, maximum epoch variance, and generic discontinuities for EEG spatial features to detect horizontal or vertical eye blinking artifacts from independent-components (ICs) \cite{hyvarinen2000independent}.  The resulting EEG artifact-free trials were baseline corrected -200 ms, prior to  stimuli onset (0ms) using a linear detrending described in \cite{torres2021facial}, then re-segmented for feature extraction to 0-1500ms. 
\subsection{ZCA whitening transformation}
The Zero Components Analysis (ZCA) is a whitening transformation used to normalize the images amplitude using a Zero Phase Mahalanobis Distance criterion \cite{coates2012learning}, without changing the correlation between the feature domains in the resulting covariance matrix.

The artifact-free input EEG image is denoted as $x$ in the following analyses. The covariance matrix associated to the input $x$, denoted as $S_{x}$, is calculated following $S_{x}=VDV^{T}$ where V is the eigen-vectors matrix of $x$ and $D$ is the diagonal matrix to construct the eigenvalue decomposition of $x$. Thus, the new whitened-image $X_{zca}$ is calculated following Equation 1 controlling the level of output contrast using $\epsilon_{zca}$. For our specific facial emotion decoding pipeline we use a low contrast of $\epsilon_{zca}=0.01$.
\begin{equation}
X_{zca}=\frac{VV^{T}x}{\sqrt{D+\epsilon_{zca}I}}
\label{zca}
\end{equation}
 The resulting $X_{zca}$ has the same size of input $x$. The new image $X_{zca}$ represents a non-rotated (i.e., zero-phase feature-space) whitened image to efficiently feed  (CNN)-based pipelines \cite{lee2018effectiveness,huang2018adversarially}. We used this new ZCA image $X_{zca}$ for training the CNN classifier on a single-trial or image level. Following this approach we could obtain a better separability than only using $x$ \cite{torres2021facial}.
\vspace{-0.2cm}
\subsection{CNN architecture and  training}
The complete pipeline for the proposed EEG-based facial emotion decoding on individuals with and without ASD we used here is described in \cite{torres2021facial}, and it is described graphically in the supplementary material - Figure S.1. In this subsection we will describe the CNN architecture including: the classifier convolutional-pooling (conv-pool) blocks, parameters, dimensions,  and training methods including: initialization, learning rates, and stopping criteria.
\begin{table*}[!htb]
  \scriptsize
  \captionsetup{font=scriptsize}
  \hspace*{-0.9cm}
  \centering
  \caption{\scriptsize{Mean and std for Face Emotion Recogniton (FER) or human performance, and  CNN (machine) performances metrics for the eighty-eight participants on this study . The results are computed averaging the Accuracy (Acc), precision (Pre), Recall (Re), and F1 score (F1) from all the confusion matrices constructed per participant. For ASD group comparing the metrics across FER and CNN modalities we found always significant differences $p<0.05$*.}
}
    \begin{tabular}{|p{1.6cm}|p{1.6cm}|p{1.6cm}|p{1.6cm}|p{1.6cm}|p{1.6cm}|p{1.6cm}|p{1.6cm}|p{1.6cm}|}
    \hline
    {\textbf{\tiny{Metrics/Groups}}} & \multicolumn{4}{p{2.4cm}|}{\textbf{FER}} & \multicolumn{4}{p{2.4cm}|}{\textbf{CNN}} \\
\cline{2-9}    \multicolumn{1}{|p{1.4cm}|}{} & \textbf{Acc} & \textbf{Pre} & \textbf{Re} & \textbf{F1} & \textbf{Acc} & \textbf{Pre} & \textbf{Re} & \textbf{F1} \\
    \hline
    \textbf{TD} & 0.815$\pm$0.083 & 0.808$\pm$0.079 & 0.802$\pm$0.077 & 0.807$\pm$0.079 & 0.860$\pm$0.213 & 0.864$\pm$0.201 & 0.860$\pm$0.204 & 0.862$\pm$0.202 \\
    \hline
    \textbf{ASD*} & 0.776$\pm$0.093 & 0.774$\pm$0.089 & 0.768$\pm$0.088 & 0.771$\pm$0.088 & 0.934$\pm$0.134 & 0.935$\pm$0.132 & 0.933$\pm$0.134 & 0.934$\pm$0.132 \\
    \hline
    \end{tabular}%
  \label{tab1}%
\end{table*}%

Our motivation for using the (CNN)-based architecture was based on previous studies which used three normalized conv-pool blocks connected to a fully-connected (FC) layer and a final decision layer \cite{torres2021facial,schirrmeister2017deep}. These type-of networks are suitable CNN architectures for the specific amount of trials we include in this study and to avoid overfitting effects. As in Schirrmeister et al. \cite{schirrmeister2017deep}, we constructed our CNN with three conv-pool blocks - going from high-to-low kernel dimensionality, and from low-to-high filters per layer. Specifically, we set the kernel dimensions per layer based on the size of $X_{zca}$, in a rectangular 30 channels $\times$ 752 time-points image, to make the conv-pool blocks more rectangular than the typical CNN architectures used for image categorization \cite{krizhevsky2012imagenet,lee2014classifying}. 

The first conv-pool block was composed of a convolutional layer with a kernel-size of $100\times10$ and 32 filters, and a subsequent max-pooling layer with a size of $5\times2$ connected to an amplitude normalization layer. The second conv-pool block was composed of a convolutional layer with a kernel-size of $20\times5$, and a max-pooling layer with a size of $2\times2$ units connected to a second amplitude normalization layer. A third conv-pool block was composed of a conv-layer with a size of $10\times2$, and a max-pooling layer with a size of $2\times2$ and 128 filters. No batch-level normalization was used. Each conv-pool block had a stride factor of 2 and non-zero-padding. Thus, the output size was half the size of the x and y dimensions - without adding any extra zeros in the image edges. The third max-pooling layer on the last conv-pool block was connected to a dense fully-connected (FC) layer with 1024 sigmoid units, and this layer was connected to a softmax layer for computing the four classes probabilities associated with a particular emotion (i.e., happy, sad, anger, and fear). This last conv-pool block did not have a normalization layer connected before the softmax layer.

The training method was based on the Adam optimization \cite{kingma2014adam}. We set an initial learning-rate equaling 0.00001 with a linear weight decay of 0.000001 per iteration. To assure a faster convergence in the training process. we used the Glorot’s initializer for the kernel weights and for all the convolutional and max-pool layers \cite{glorot2011deep}. For the biases initialization, we use an uniform random distribution across all the layers with $\mu=0$ and $\sigma=0.1$. No random or heuristic search was used to set the initial learning rate or the decay rates. We used a dropout layer with $p=0.25$ applied to the FC layer. All the conv-pool activation-functions were Rectified Linear Unit (ReLU) and trained with 4 size mini-batches - changing the training indexes randomly on each iteration. A maximum of 500 iterations was set as part of the training process. We used the early-stopping  criterion described in \cite{schirrmeister2017deep} and 72.5\% of trials across all participants fell into this early stop criterion requiring less than 200 iterations - 73.31\% TD and 74.65\% ASD.

The (CNN)-based pipeline performance was evaluated using a Leave-One-Trial-Out (LOTO) cross-validation for each participant. This means that the performance was measured iterating intra-subjectly across all the 48 EEG trials, per participant, using 47 for train and 1 for testing. The Accuracy (Acc), precision (Pre), Recall (Re), and F1 scores reported on Table \ref{zca} are calculated using a wrapped (macro) confusion matrix calculated for each participant obtained in the LOTO cross-validation. We used and reported the same metrics in our previous study \cite{torres2021facial}.
   The evaluation of the metrics for the human behavioral performance on the  FER task consists of the same approach explained above - assuming each participant has an equal level of entropy as observed in the 47 trials used for training the CNN. These performance metrics using the LOTO cross validation provide a personalized neural representation for facial emotion decoding in individuals with and without ASD \cite{torres2021facial}.
\begin{figure*}
\centering
\captionsetup{font=scriptsize}
\vspace{-1cm}
\hspace*{-1.1cm}
\subfloat[\small{Smooth-Grad Squared TD}]{
   \includegraphics[width=8.5cm, height=4.0cm]{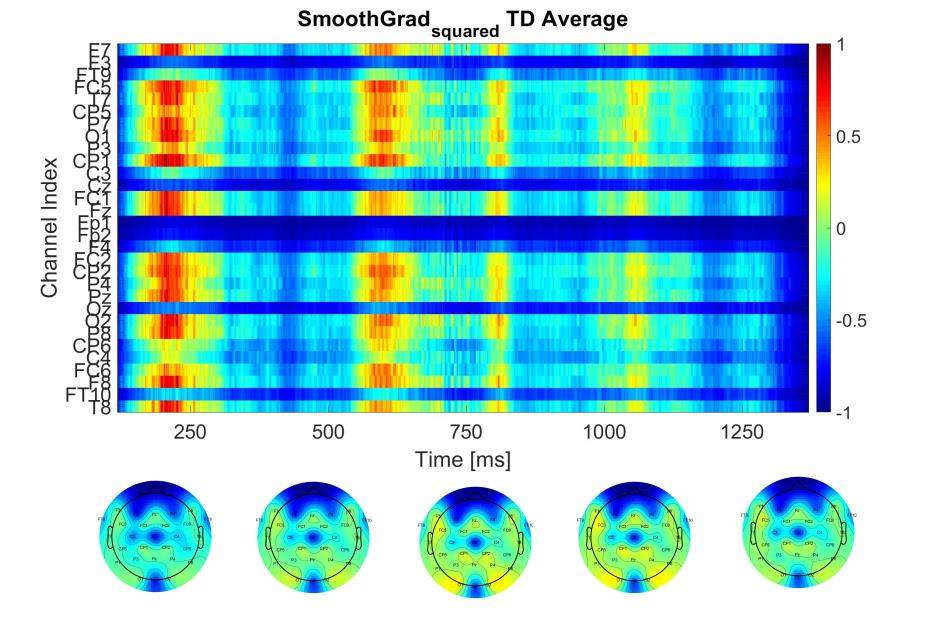}
   \label{smooth_grad_squared_TD}
} 
\subfloat[\small{Smooth-Grad Squared ASD}]{
   \includegraphics[width=8.5cm, height=4.0cm]{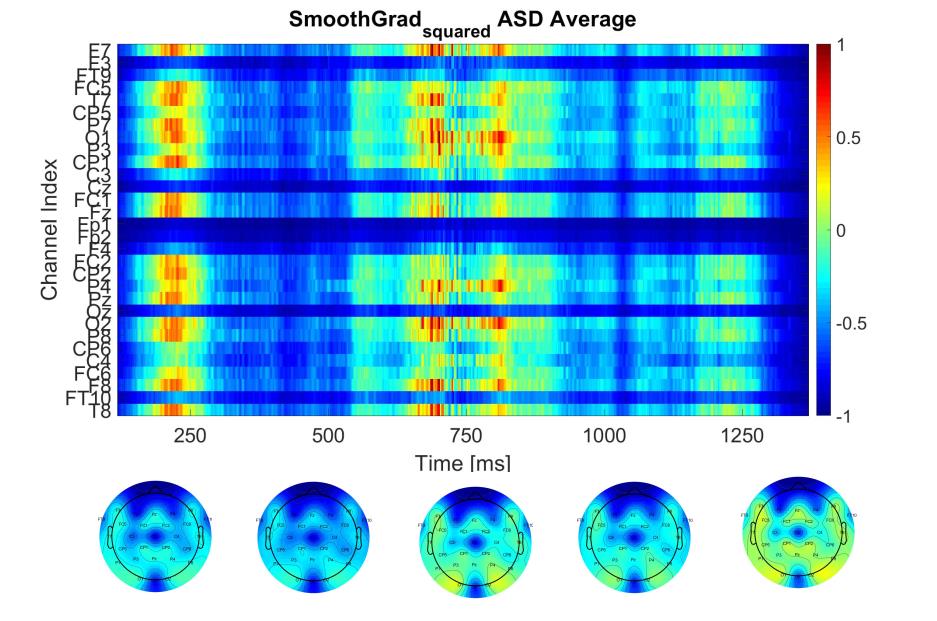}
   \label{smooth_grad_squared_ASD}
}
\\
\hspace*{-1.1cm}
\subfloat[\small{PatternNet TD}]{
   \includegraphics[width=8.5cm, height=4.0cm]{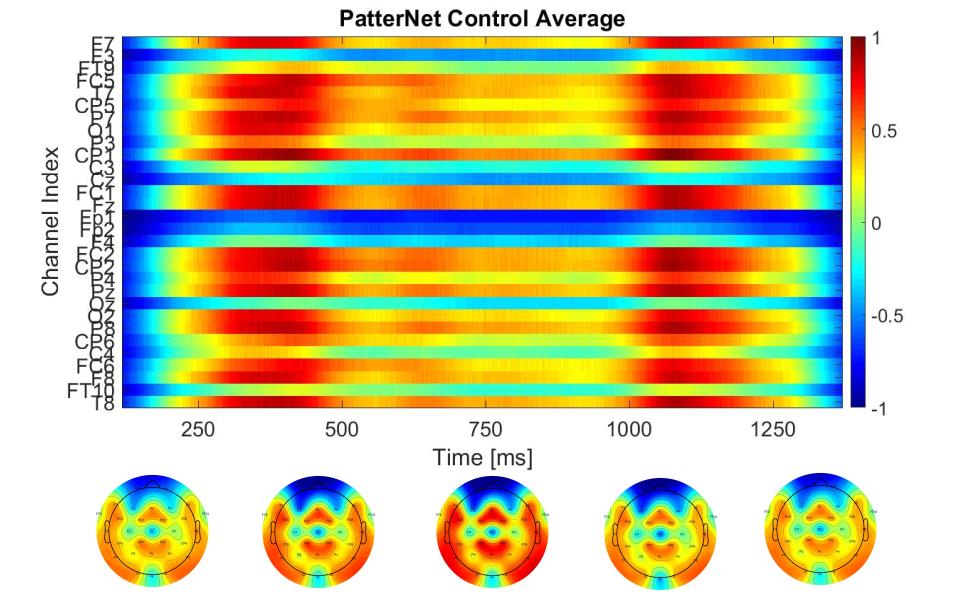}
   \label{PatternNet_TD}
} 
\subfloat[\small{PatternNet ASD}]{
   \includegraphics[width=8.5cm, height=4.0cm]{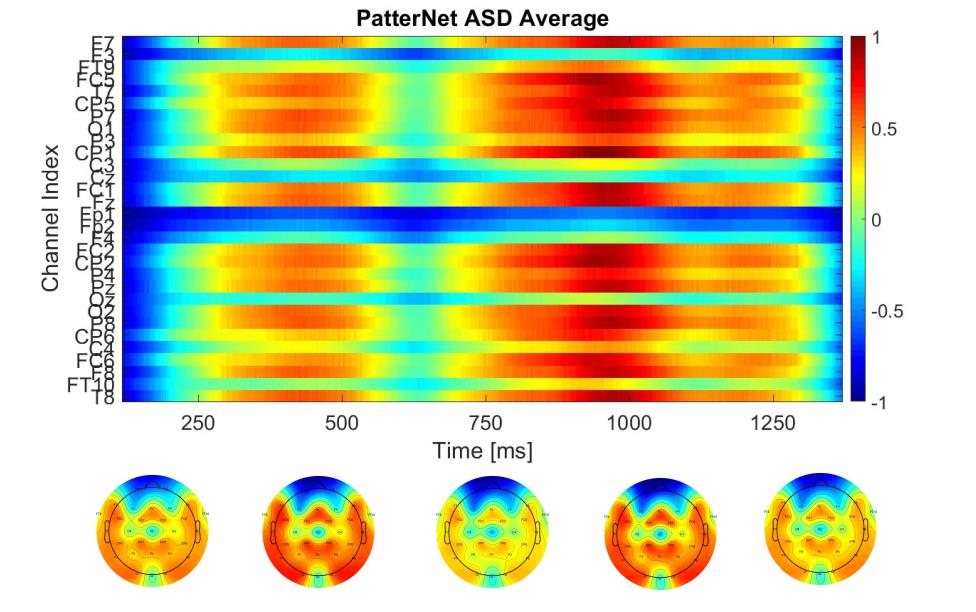}
   \label{PatternNet_ASD}
}
\\
\hspace*{-1.1cm}
\subfloat[\small{Pattern-Attribution TD}]{
   \includegraphics[width=8.5cm, height=4.0cm]{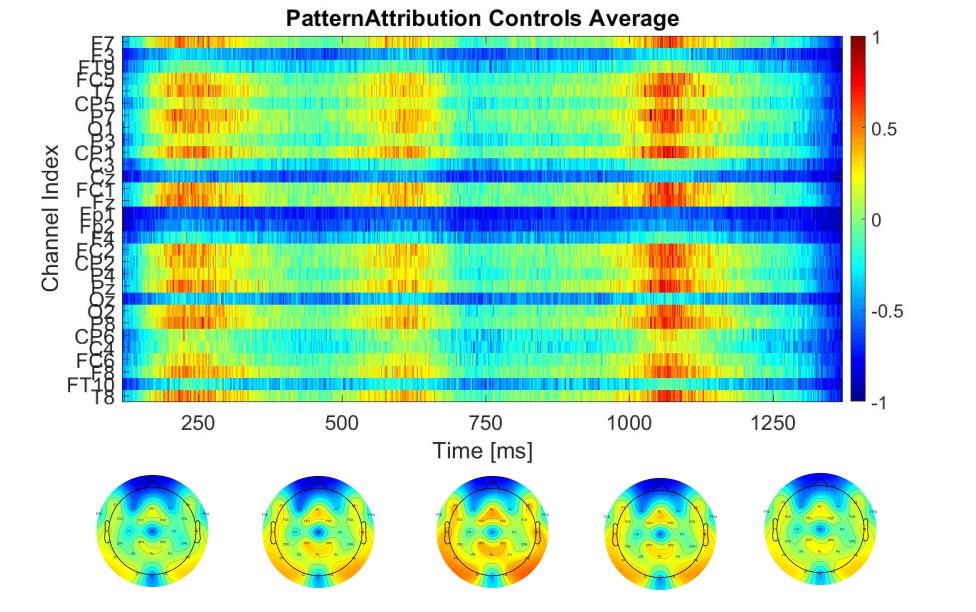}
   \label{patternattr_TD}
} 
\subfloat[\small{Pattern-Attribution ASD}]{
   \includegraphics[width=8.5cm, height=4.0cm]{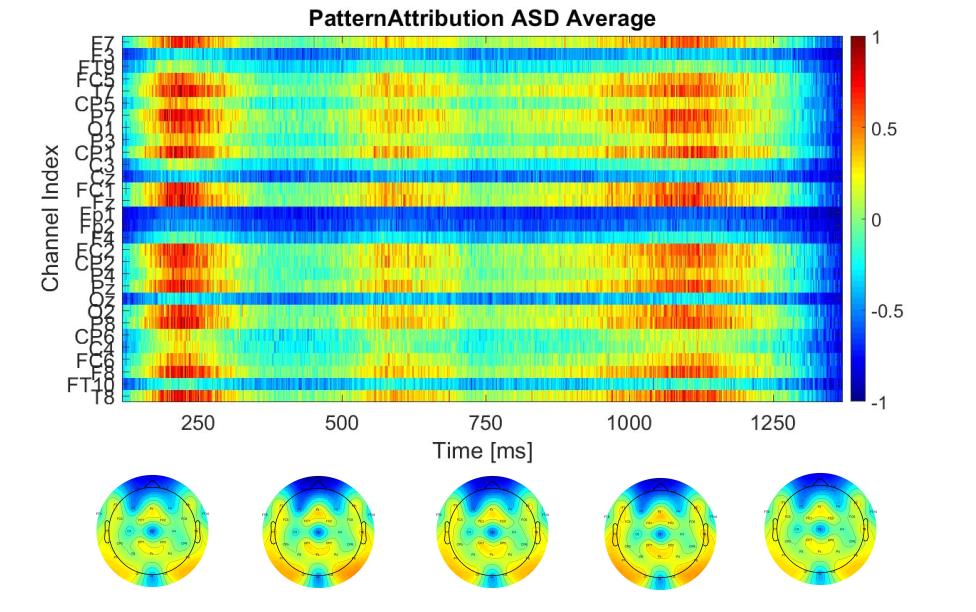}
   \label{patternattr_ASD}
}
\\
\hspace*{-1.1cm}
\subfloat[\small{LRP B TD}]{
   \includegraphics[width=8.5cm, height=4.0cm]{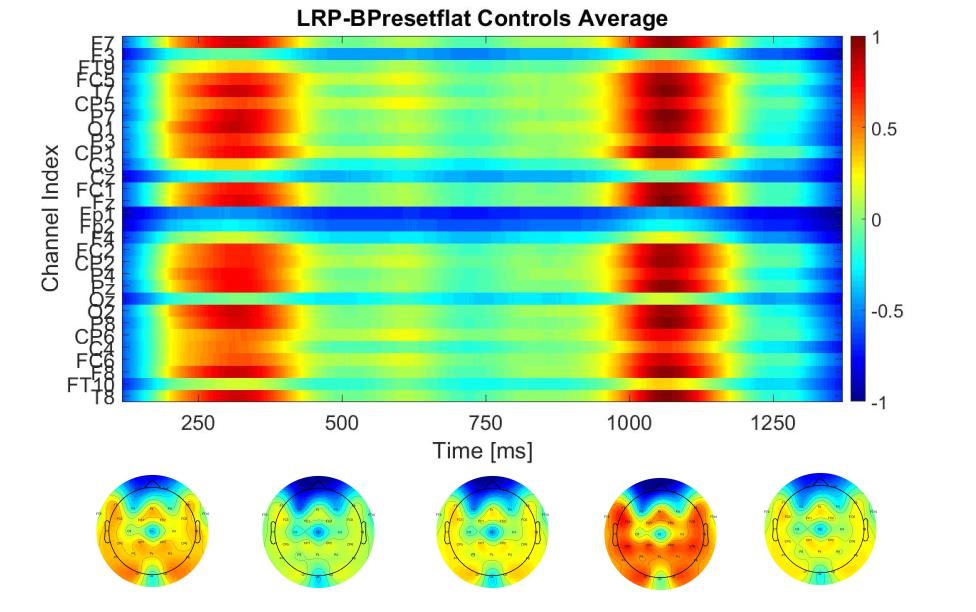}
   \label{LRP_B_TD}
} 
\subfloat[\small{LRP B ASD}]{
   \includegraphics[width=8.5cm, height=3.9cm]{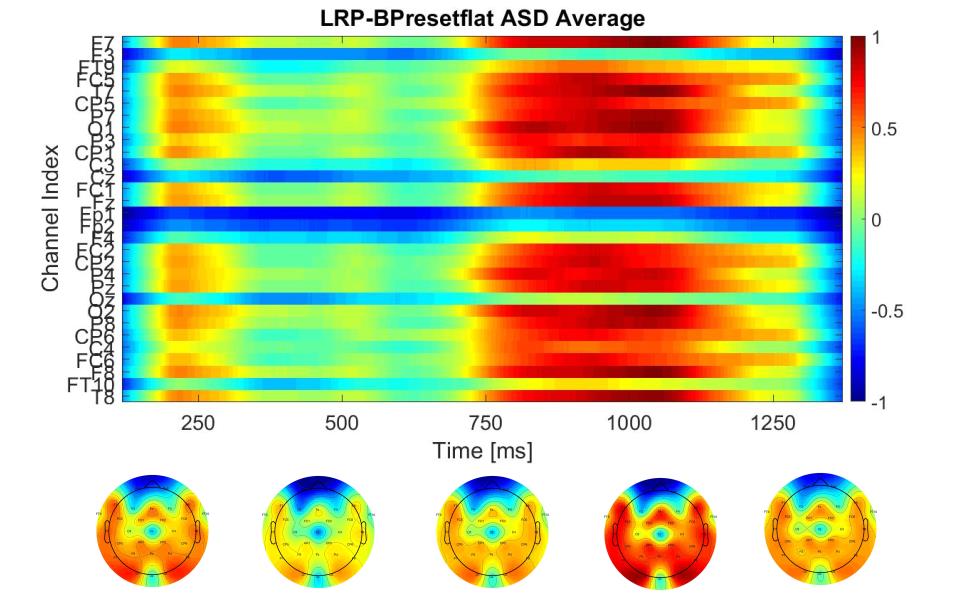}
   \label{LRP_B_ASD}
}
\caption{\scriptsize{Average relevance-maps for Smooth-Grad Squared, PatternNet, Pattern-Attribution, and LRP-B flat preset. The relevance-maps for TD are shown on the left, and for ASD on the right. These relevance maps are normalized between [-$1,1$] and coloured using the jet colormap being the more relevant values on dark-red, and the more unrelevant on darker-blue.}}
\label{fig:rel_maps}
\end{figure*}
\vspace{-0.2cm}
\subsection{XAI methods}
We used four XAI methods for estimating feature-importance levels on the trained CNN. We used the Smooth-Grad \cite{smilkov2017smoothgrad} method as a baseline and three other XAI saliency-maps: Smooth-Grad Squared \cite{adebayo2018sanity} - a simple numerical variation of Smooth-Grad, PatternNet, Pattern-Attribution \cite{kindermans2017patternnet}, and the Layer-Wise Relevance Propagation (LRP) \cite{binder2016layer}. These XAI  methods are included in the \emph{iNNvestigate} software package \cite{alber2019innvestigate}, a Python module we used for evaluating each XAI method proposed in this study. 

For the subsequent analyses we define a feature relevance-map based on the LRP model \cite{montavon2018methods,samek2020toward},  and for all the XAI methods analyzed here as $R_{q}^1$. These relevance values are normalized in amplitude, limiting the propagated relevance between [-$1,1$]. Thus, the $R_{q}^1$ $\geq$0 values are considered \emph{relevant} - or a positive contribution to successful facial emotion decoding. All the features associated with positive values represent a hit or an accurate CNN facial emotion decoding. On the other hand, the $R_{q}^1$ $<$0 values are considered \emph{irrelevant} or a negative contribution for a successful facial emotion decoding. 

To obtain a final average relevance measure for each facial emotion, we average the resulting relevance map for each iteration - across the iterations of the LOTO -per subject- cross-validation \cite{torres2021facial}. Figure \ref{fig:rel_maps} shows the average relevance maps for TD and ASD,  and for all the XAI methods. In the following subsections, we describe the detailed models of each XAI method included in this study for statistical evaluation and ROAR.
\subsubsection{Smooth-Grad and Smooth-Grad Squared}
Smooth-Grad is a XAI  method proposed by Smilkov et al. \cite{smilkov2017smoothgrad}.  In the basic form they start out by computing the gradient of logit (unit) i w.r.t. to the input, denoted as $g(x)=\frac{\partial f(x)_{i}}{\partial x}$. Typically this results in noise saliency-maps which can be mitigated by averaging the gradients of multiple noisy versions of the input, thereby improving the signal to noise ratio.
\begin{equation}
\hat{g}(x)=\frac{1}{N}\sum_{i=1}^{N}g(x+\mathcal{N}\left(0,\sigma^{2}I\right))
\label{smoothgrad}
\end{equation}
The hyper parameters for this method are the number of samples N we use, the mean – typically zero -- and variance 2 of the Gaussian noise. From Equation \ref{smoothgrad}, it should be clear that this method can in principle be applied to any XAI method and not just the gradients. Recent studies \cite{hooker2018evaluating,kindermans2017reliability} used a variation called Smooth-Grad squared. Here the gradients are squared before averaging as shown in Equation \ref{smoothgrad_squared}.
\begin{equation}
\hat{g}(x)=\frac{1}{N}\sum_{i=1}^{N}g(x+\mathcal{N}\left(0,\sigma^{2}I\right))^2
\label{smoothgrad_squared}
\end{equation}
Previously, this approach performed better than the original Smooth-Grad implementation for vision tasks \cite{adebayo2018sanity,hooker2018evaluating}. The relevance maps for TD and ASD groups calculated using the Smooth-Grad and the Smooth-Grad Squared  are shown from subfigure \ref{smooth_grad_squared_TD} to \ref{smooth_grad_squared_ASD} - in Figure \ref{fig:rel_maps}.  We consider SmoothGrad as a baseline because of its simplicity and its broad usage on other XAI studies \cite{kindermans2017reliability}.
\vspace{-0.2cm}
\subsubsection{Layer-Wise Relevance Propagation}
Layer-Wise Relevance Propagation (LRP) \cite{binder2016layer,bach2015pixel} is a family of XAI methods based on the Lebesgue energy-conservation law \cite{szepessy1989existence}. The quantity observed at the logit $i$ is seen as the relevance for a certain class and in each layer of the network we assume that the same amount of relevance is present for this class. Equation \ref{LRP_lebesgue} makes this explicit for a multiple-layer network. The relevance observed at the logit is equal to $f(x)_{i}$. This is then distributed towards the input in a layerwise manner such that relevance is preserved through the layers. The relevance at neuron $d$ in layer $l$  is denoted as $R_{d}^l$ in LRP.
\begin{equation}
\small
f(x)=\sum_{q}R_{q}^{1}=\sum_{d \in (l+1)}{R_{i,j}}_{d}^{l+1}=\sum_{d \in (l)}{R_{i,j}}_{d}^{l}= \hdots = \sum_{d}{R_{i,j}}_{d}^{L}
\label{LRP_lebesgue}
\end{equation}
LRP methods have variations of this process that are optimized for specific tasks.  In this study, we used LRP preset B (LRP-B) as implemented in the iNNvestigate toolbox. This LRP configuration makes use of the epsilon LRP rule for dense layers and the alpha-beta rule for convolutional layers. 

The epsilon rule is based on the z-rule where the relevance is proportional to the weight multiplied neuron contribution denoted as $z_{ij}=x_{i}^{l}\omega_{ij}^{(l,l+1)}$. These contributions are then normalized as seen below.
\begin{equation}
R_{i}^{l}=\sum_{j} \frac{z_{ij}}{\sum_{i}z_{i,j}}R_{j}^{l+1}=\sum_{j} \frac{x_{i}^{l+1}\omega_{ij}^{(l,l+1)}}{\sum_{i}x_{i}^{l}\omega_{ij}^{(l,l+1)}+b_{j}}R_{j}^{l+1}
\label{LRP_eq}
\end{equation}
This normalization can become problematic if the denumerator/denominator is (close to) zero. In this case, a very small value epsilon is added (or subtracted to keep the sign constant) to ensure numerical stability. For the convolutional  layers the $\alpha$-$\beta$ rule is used.
\begin{equation}
R_{i}^{l}=\sum_{j}\left[\alpha\frac{z_{ij}^{+}}{\sum_{j}z_{ij}^{+}}+\beta\frac{z_{ij}^{-}}{\sum_{j}z_{ij}^{-}}\right]
\label{LRP_alpha_beta}
\end{equation}
 Using this approach with the alpha-beta rule includes parameters $\alpha$ and $\beta$ that controls how much weight is given to the positive relevance components ($z_{ij}^{+}$), and negative relevance components ($z_{ij}^{-}$). While the iNNvestigate package \cite{alber2019innvestigate} includes many variants of the LRP rules as presets, we use preset B  as described above with $\alpha=2$ and $\beta=1$. Preset A with $\alpha=1$ and $\beta=0$. was also evaluated, but was less significant that preset B on our data and therefore is not included \cite{torres2021facial}.
\subsubsection{PatternNet and Pattern-Attribution}
PatternNet and Pattern-Attribution XAI salience methods have previously been described by P.J Kindermans et al. \cite{kindermans2017patternnet}. These methods can be seen as an extension of the LRP or Deep-Taylor Decomposition framework. PatternNet and Pattern-Attribution  consider which parts of the input a neuron is invariant to and which parts it is trying to detect. By doing this, the explanation of a single artificial neuron without the non-linearity is consistent with the explanation of multivariate linear models in neuroimaging \cite{winkler2015influence}. PatterNet and Pattern-Attribution do still rely on layerwise propagation from LRP to combine individual neuron-wise explanations to whole network explanations. 

In Pattern-Attribution and PatternNet XAI methods the input $x$ to a neuron is assumed to be composed of an informative signal component and a non-informative distractor component from the output of that neuron $y$.
\begin{equation}
x=a_{s}y + a_{d}\epsilon
\label{patternetpattern}
\end{equation}
Here $\epsilon$ can be seen as a noise source that minimizes the distractor so it is essentially ignored. PatternNet and Patter-Attribution operate under the assumption that the linear model explanation should not change if  is zero. Therefore, the estimation of $a_{s}$ is necessary to propagate according to the signal component.  PatternNet and Pattern-Attribution propose Equation \ref{patternetpattern} to find a set of new filter weights $\hat{\omega}$ that are separated as much as possible from the network weights $\omega$- in order to maximize the signal component across the layers of the trained CNN following $y^{l}=\hat{\omega}^{T}x^{l}y^{l}$ is the output of the layer $l$ and $x^{l}$ the corresponding input associated with the filter weights $\hat{\omega}$. 

For PatternNet and Pattern-Attribution the calculation of the distractor $d$ assumes the following equivalences: $y^{l}=\hat{\omega}^{T}x^{l}$ and $\hat{\omega}^{T}d=0$. We obtain this latter equivalence deriving Equation \ref{patternetpattern} in terms of $\hat{\omega}$ and optimizing. Thus, to compute a new signal estimator that isolates the network noise from  we can define $S_{a}(x)=a_{d}\hat{\omega}^{T}x$. This new estimator increases the correlation between $x$ and $s$ in a new term denoted as $\rho$. This correlation can be calculated with the estimation of the input distribution of $x$ - denoted as u on each subsequent layer, and considering  $\hat{\omega}^{T}d=0$ during the relevance propagation across all those layers after training. $\rho$ is then defined in Equation \ref{optim} - assuming all the noise will be reflected in the estimated variance $\sigma_{u,d}=\sigma_{y}$.
\begin{equation}
\rho(S(x)) =1-\texttt{max}\left(\frac{u^{T}cov(d,y)}{\sqrt{\sigma_{u,d}\sigma_{y}}}\right)
\label{optim}
\end{equation}
 In order to assure that $\hat{\omega}^{T}d=0$ will occur during the relevance propagation, the covariance between the distractor d and the layer output y  must be zero or as close as possible to zero $cov(d,y)$. In this manner, the new signal estimator $S_{a}(x)$, or for simplicity reasons $a$, can be obtained assuming that $cov(x,y)=cov(S_{a}(x),y)$. Thus, assuming that the learning rates are independent of the interaction between $x$ and $y$ on each layer,  we can obtain $cov(x,y)=acov(y,y)$ and the definitive signal estimator $a$ is defined in Equation \ref{estimator}. \begin{equation}
a = \frac{cov(x,y)}{\sigma_{y}} = \frac{\left[E_{+}[xy]-E_{+}(x)E_{+}(y)\right]}{\left[\omega^{T}E_{+}[xy]-\omega^{T}E_{+}(x)E_{+}(y)\right]}
\label{estimator}
\end{equation}
$E$ is the expected value for any single or joint variable $xy$ described in Equation \ref{estimator}. The propagation of the estimator $a$ is the main difference between PatternNet and Pattern-Attribution. Both methods use the same model described on Equation \ref{estimator} to calculate the estimator per layer. However,  PatternNet uses a similar propagation as the LRP Deep-Taylor model \cite{binder2016layer,montavon2018methods} based on the "message passing" modality - without propagating the estimator using information from $\hat{\omega}$. 
Pattern-Attribution method uses the numerical incidence of the filter weights $\hat{\omega}^T$ which are estimated through $\omega$. This method propagates a part of $\hat{\omega}$ using a product between the filter weights and the signal estimator $\hat{\omega}^Ta$, instead of only $a$ \cite{kindermans2017patternnet}. This last consideration of filter weights supports a more noisy relevance map calculated from Pattern-Attribution as we can see in Figures \ref{PatternNet_TD} and \ref{PatternNet_ASD}.
   
The top part of each subfigure in Figure \ref{fig:rel_maps} are the Average saliency-maps with the channel indexes on the y-axis, and the time-points in ms on the x-axis. The bottom part of each subfigure shows five topo-plots representing the average relevance on five different time ranges, such as 0-500, 250-750, 500-1000, 750-1250, and 1000-1500 ms.
\subsection{Remove-And-Retrain (ROAR) - Certainty Analysis}
For evaluating the certainty and reliability of the XAI methods, we used the RemOve-And-RetrAin (ROAR) methodology \cite{hooker2018evaluating}. ROAR uses the average relevance map $R_{s}^1$ for each participant group (i.e., TD and ASD) for each facial emotion as a feature importance  indicator. ROAR weights the feature-importance directly from the input feature-space using the relevance values calculated from each XAI method and sorts the values from high-to-low relevance.
The ROAR pipeline is reported in Figure S.2 in supplementary material. In ROAR, the average XAI relevance map, per group, $R_{s}^1$ determines whether  features are included in each new training process. Features are suppressed in order of high-to-low relevance. Feature removal is done based on an element-wise product between the average posthoc relevance map  obtained from an XAI method and the input EEG image. The product dictates what features will be removed or accepted for  the CNN re-training based on a relevance threshold.

Previous studies have used ROAR to assess the level of certainty of multiple XAI methods in a quantitative way [64,65]. These studies compare the level of accuracy detriment associated with the relevant features removal using  random relevance patterns as a baseline. We used this same approach to evaluate and compare XAI methods using ROAR and random relevance baselines within each group. To do this, we set various relevance thresholds as a pixel/feature-removal rate $r$ ($r$=0.1, 0.2, 0.5, 0.7, and 0.9) to generate a binarized-mask, $R_{b}$, from the average relevance-map $R_{s}^1$ using Equation \ref{r_rate}.

\begin{equation}
R_{b}=\begin{cases} 
      1 & R_{s} \leq r \\
      0 & R_{s}>r
   \end{cases}
\label{r_rate}
\end{equation}

The relevance $R_{s}^1$ is averaged across facial emotions  $R_{s}^1=(R_{happy}+R_{sad}+R_{angry}+R_{fear})/4$ for normalization.  The resulting normalized binary-mask dictates which pixels/features are admitted into the re-training. These binary masks have the same channels and time-points indexes and size of $X_{zca}$. Different binary masks used in this analysis are reported and illustrated in the supplementary material in Figures S.3 and S.4 including different values of $r$. 

In the following subsection, we will introduce the random baselines we used  to compare the final ROAR metrics, and check if the re-training performances are more interpretable using the XAI methods or the random patterns in the feature removal.
\subsubsection{ROAR baselines}
We include two basic distribution-based binary masks as random baselines for the ROAR evaluation. The first, most uninformed, baseline we used is a common random baseline based on an uniform distribution with a pixel/feature removal set to $r$=0.5. 
   
The second baseline we used evaluated time-domain feature relevance per channel. In this baseline, two  types of random slices are generated. The first slice is a random pattern based on a 47 $\times$ 1 slice covering 47 time-points and  a single channel. This pattern allows all features in ROAR, while occluding time-domain slices with a size of 47 time-points $\times$ 1 channel. This retains a cohesive set of slices of approximately 20ms around important positive or negative EEG event-related potentials (ERPs) evoked during FER, such as N1, P3, and Late-Positive Potential  (LPP) \cite{webb2017face,cohen2017does}. The second slicing baseline is identical to the first, but does not sort the 47$\times$1 slices randomly. Instead, it sorts relevance values based on a XAI method to dictate which feature will be removed. This pattern will be referred to as a method-related slice baseline in the following analyses. Binary-masks are reported in the supplementary material in Figure S.5. 
As a gold-standard, we expect that the feature removal associated with the more reliable XAI saliency-maps reduces more accuracy than the random baselines. We also expect that the accuracy detriment will be more plausible for higher $r$ values in comparison with lower $r$ values, thus assuring that those removed features are truly relevant.
\section{Results}
This results section is divided in three subsections 1) human behavioral FER and CNN performance results for both the TD and ASD groups 2) ROAR  results comparing XAI salience methods relative to random baselines and each other within groups for  interpretability of   FER encoding, and 3) group differences in XAI relevance maps for each facial emotion or class.
\vspace{-0.2cm}
\subsection{Performances - FER and CNN}
Table \ref{tab1} includes the Accuracy (Acc), precision (Pre), Recall (Re), and F1 scores for human behavioral FER and CNN performances. The metrics are formally described in \cite{powers2011evaluation}.
Using one-way ANOVA, we found significant differences in Acc (F(1,87)=10.43, p=0.00144), pre (F(1,87)=6.31, p=0.0301), Re (F(1,87)=9.35, p=0.00561), and F1 (F(1,87)=8.66, p=0.0232), between FER and CNN metrics. For all metrics, CNN was more accurate than FER.
\subsection{XAI methods - Statistical comparisons}
The five time ranges mentioned above are used for adjusting the p-values of the statistical comparison reported below using a Bonferroni-Holm correction \cite{abdi2010holm}.
\vspace{-0.2cm}
\subsubsection{XAI saliency-maps comparison}
Visual inspection of XAI saliency-maps in Figure \ref{fig:rel_maps}, shows similarities in the relevance distribution within the  five time-ranges mentioned above. To test these statistical similarities, we used the Kolmogorov-Smirnov test (KS-test) \cite{banerjee2018kolmogorov}. Comparing the relevances obtained in the Smooth-Grad method and Smooth-Grad Squared method we found similarities across the five time ranges h=1, p$\leq$ 0.001**.

Similarities were also observed in late-time ranges, such as 750-1250 and 1000-1500 ms, between PatternNet and LRP-B h=1, p$\leq$0.028*, and between PatternNet and Pattern-Attribution, h =1, p$\leq$0.043*.  We attributed this to a more noisy relevance pattern obtained from Pattern-Attribution after the marginal values of the weights $\hat{\omega}^T$ can also modulate the calculation of the estimator $a$ (see  section 2.7.3). There were no other statistical similarities between other XAI methods, h=0, p’s$>$0.05. All these findings are in line with our expectations after training the CNN with this type of EEG image.
\vspace{-0.2cm}
\subsubsection{Differences between TD and ASD relevance maps}
There were no significant differences in the Smooth-Grad,Smooth-Grad average, Smooth-Grad Squared, PatternNet, or Pattern-Attribution XAI salience maps for any specific facial emotion, or the average across facial emotions  between  TD and ASD groups. For Smooth-Grad we obtained F’s(1,87)$\leq$1.334, p’s$>$0.05, across all emotions or any time-range (see section XAI methods). Similar differences were obtained for Smooth-Grad Squared F(1,87)$\leq$0.129 with p$>$0.05, PatternNet F(1,87)$\leq$0.883, p’s$>$0.05, and Pattern-Attribution F(1,87)$\leq$0.222, p’s$>$0.05. 

The main significant differences found between TD and ASD XAI salience maps are observed in the LRP-B method, specifically for  negative emotions, such as Anger and Fear. This is consistent with the observable behavioral deficits in ASD performing FER. We found significant differences for Average in 0-500ms, (F(1,87)=7.889, p=0.0344), TD$>$ASD, and in 1000-1500ms, (F(1,87)=11.56, p=0.0033),  TD$<$ASD. For Sad in 750-1250ms, (F(1,87)=8.491, p=0.0141), TD$>$ASD, and in 1000-1500ms, (F(1,87)=13.54, p=0.0005), TD$>$ASD. For anger in 0-500ms, F(1,87)=10.85, p=0.0095, TD$>$ASD, and in 1000-1500ms, (F(1,87)=9.667, p=0.0102), TD$<$ASD. For Fear in 0-500ms, (F(1,87)=23.47, p=7.6E-6), TD$>$ASD, in  500-1000ms, (F(1,87)=7.193, p=0.0263), TD$<$ASD, and in [750-1250]ms, (F(1,87)=9.313, p=0.0121), TD$<$ASD. All individual F and p values for each facial emotion, group, and XAI method are reported in the supplementary material in Table S.1.
\vspace{-0.2cm}
\subsubsection{Differences between binary masks}
To evaluate the differences between binary masks obtained from all the trained TD and ASD CNNs for each facial emotion we used an One-way ANOVA and Bonferroni-Holm correction over the same five time-ranges. 
\begin{figure*}
\centering
\captionsetup{font=scriptsize}
\hspace*{-1.1cm}
\subfloat[\small{500-1000ms binary mask topo-maps}]{
   \includegraphics[trim={0 0.5mm 0 0.8mm},width=8.8cm, height=8.6cm,clip]{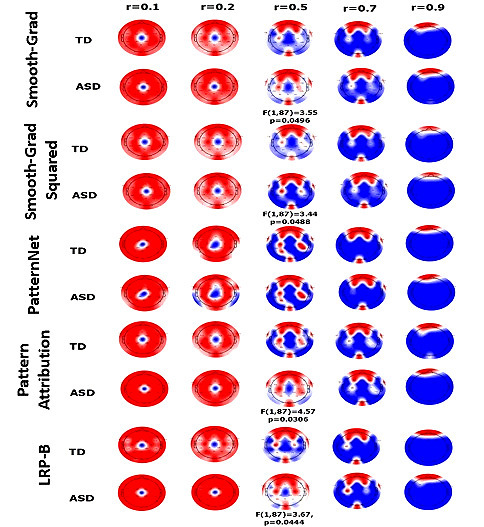}
   \label{bin_500}
} 
\subfloat[\small{750-1250ms binary mask topo-maps}]{
   \includegraphics[trim={0 0.08mm 0 0.8mm},width=8.8cm, height=8.6cm,clip]{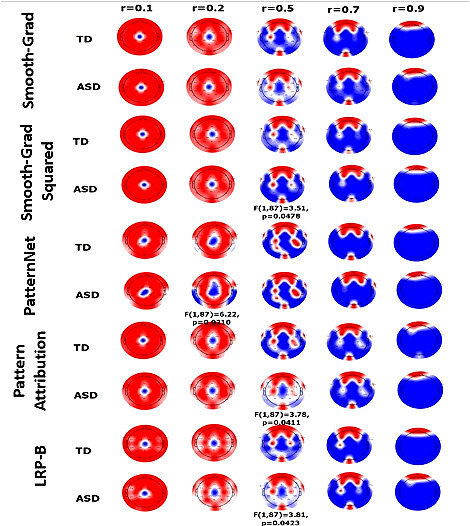}
   \label{bin_750}
}
\caption{\scriptsize{Topoplots examples for the binary masks on 500-1000ms and 750-1250ms time ranges. These topo-maps cover all the saliency methods analyzed in this study such as Smooth-Grad, Smooth-Grad Squared, PatternNet, Pattern-Attribution, and LRP-B flat preset in rows, and the $r$ values in columns. The plots are illustrated using a \emph{redblue} colormap with limits between [-$0.2,0.2$]. F and p-values are reported only for the binary-mask comparisons that are significantly different after correction.}}
\label{binary_differences}
\end{figure*}
Figure \ref{bin_500} shows the topo-plots related to the binary masks for the time range between 500-1000ms. In this time range we only found significant differences for $r$=0.5. Specifically, we found differences in Smooth-Grad (F(1,87)=3.55, p=0.050), Smooth-Grad Squared (F(1,87)=3.44, p=0.049), Pattern-Attribution (F(1,87)= 4.57, p=0.031), and LRP-B (F(1,87)=3.67, p=0.044). In PatternNet we did not find any significant difference on this early time range. This must be related to the behavioral associated with the CNN, less noisy, and coarser relevance patterns associated with PatternNet.
 
Figure \ref{bin_750} shows the topo-plots related to the binary masks for the interval between 750-1250ms. There we found differences for $r$=0.2 and PatternNet (F(1,87)=6.22, p=0.021). We also found differences for $r$=0.5 in Smooth-Grad Squared (F(1,87)=3.51, p=0.048), Pattern-Attribution (F(1,87)=3.78, p=0.041), and the LRP-B  preset (F(1,87)=3.81, p=0.043). These results suggest that in terms of the $r$ values, the binary mask patterns are consistent. This also shows that binary masks are consistent across $r$ in the late time-ranges - excepting  $r$=0.5. To assure a more consistent comparison between the binary masks of TD and ASD,  we evaluate them using the K-S test. For all the significant differences found in this analysis we can surmise that the binary masks all come from the same distribution - h=1, p$<$0.05. There were no significant differences between 0-500ms for any XAI method.
\vspace{-0.2cm}
\subsection{Accuracy detriment differences across r values}
For comparing the accuracy detriment across the $r$ values  of 0, 0.1, 0.2, 0.35, 0.5, 0.7, 0.9, and 1, we used an one-way ANOVA with a subsequent Bonferroni-Holm correction across the $r$ values. The comparison was evaluated and adjusted by grouping the accuracies obtained for each TD and ASD participant described in Table \ref{tab1}. 
\begin{figure*}
\centering
\hspace*{-1.1cm}
\captionsetup{font=scriptsize}
\subfloat[\scriptsize{TD - Saliency maps}]{
   \includegraphics[width=9.6cm, height=3.9cm]{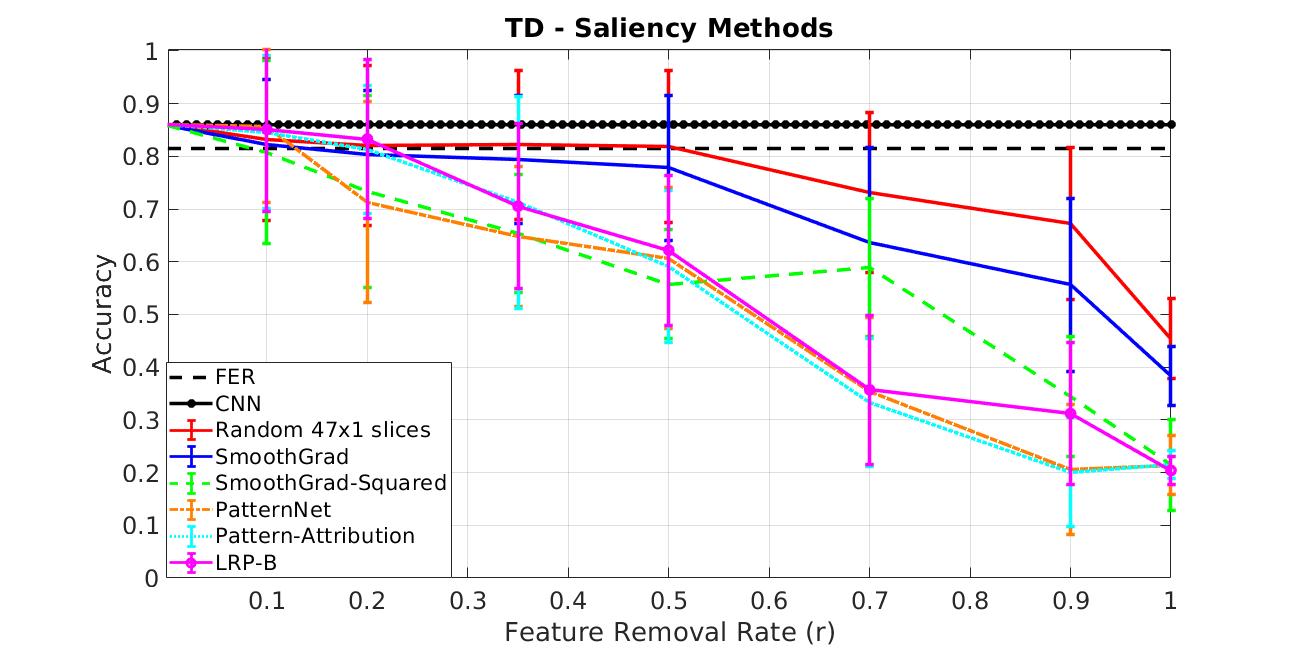}
   \label{fig_TD_all}
} 
\subfloat[\scriptsize{ASD - Saliency maps}]{
   \includegraphics[width=9.6cm, height=3.9cm]{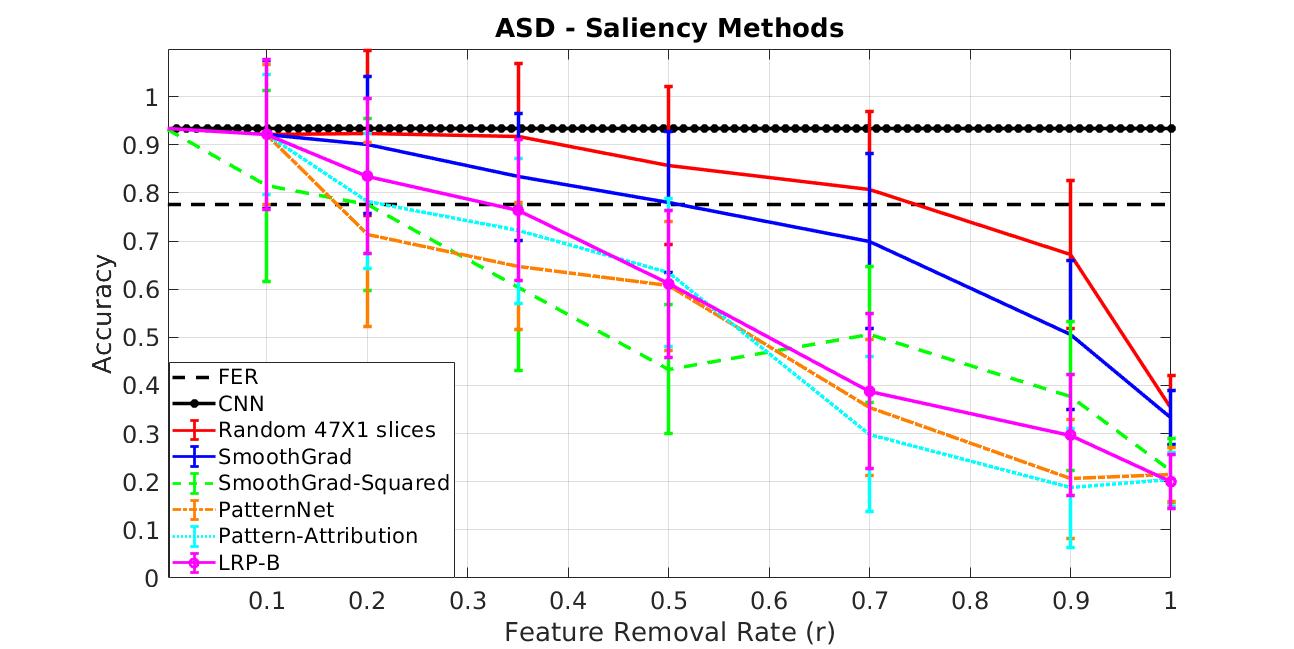}
   \label{fig_ASD_all}
}
\\
\hspace*{-1.1cm}
\subfloat[\scriptsize{TD - slices}]{
   \includegraphics[width=9.6cm, height=3.9cm]{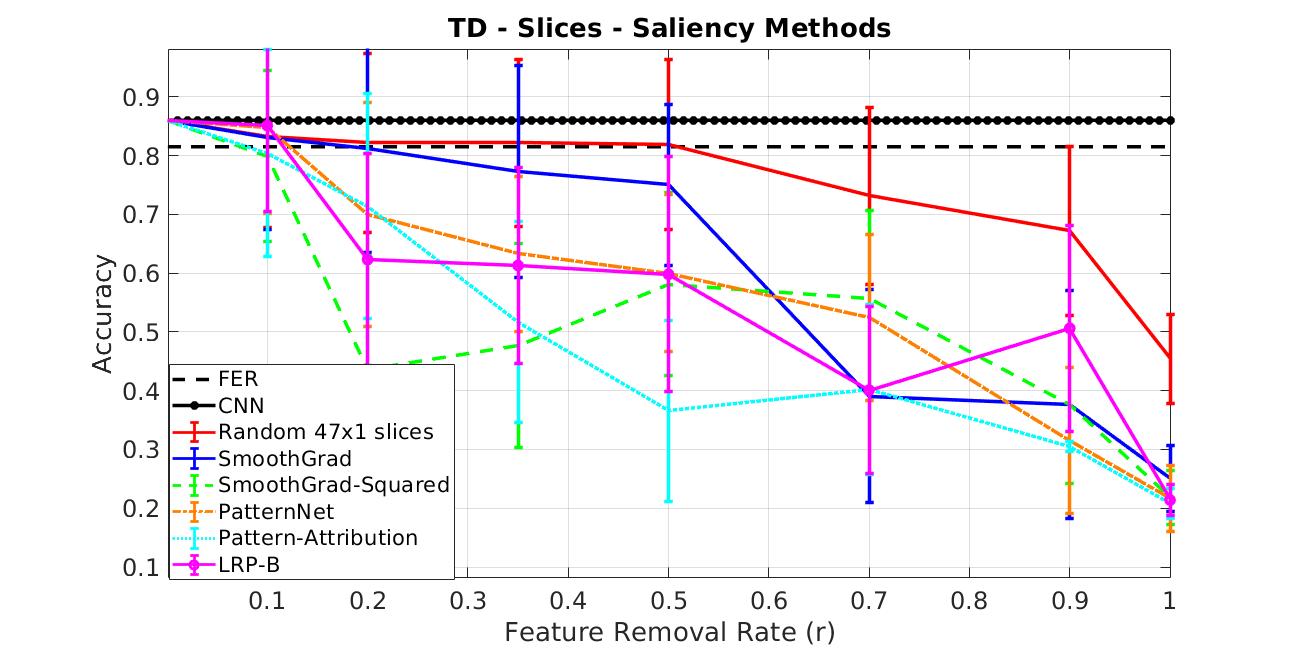}
   \label{fig_TD_slices}
} 
\subfloat[\scriptsize{ASD - slices}]{
   \includegraphics[width=9.6cm, height=3.9cm]{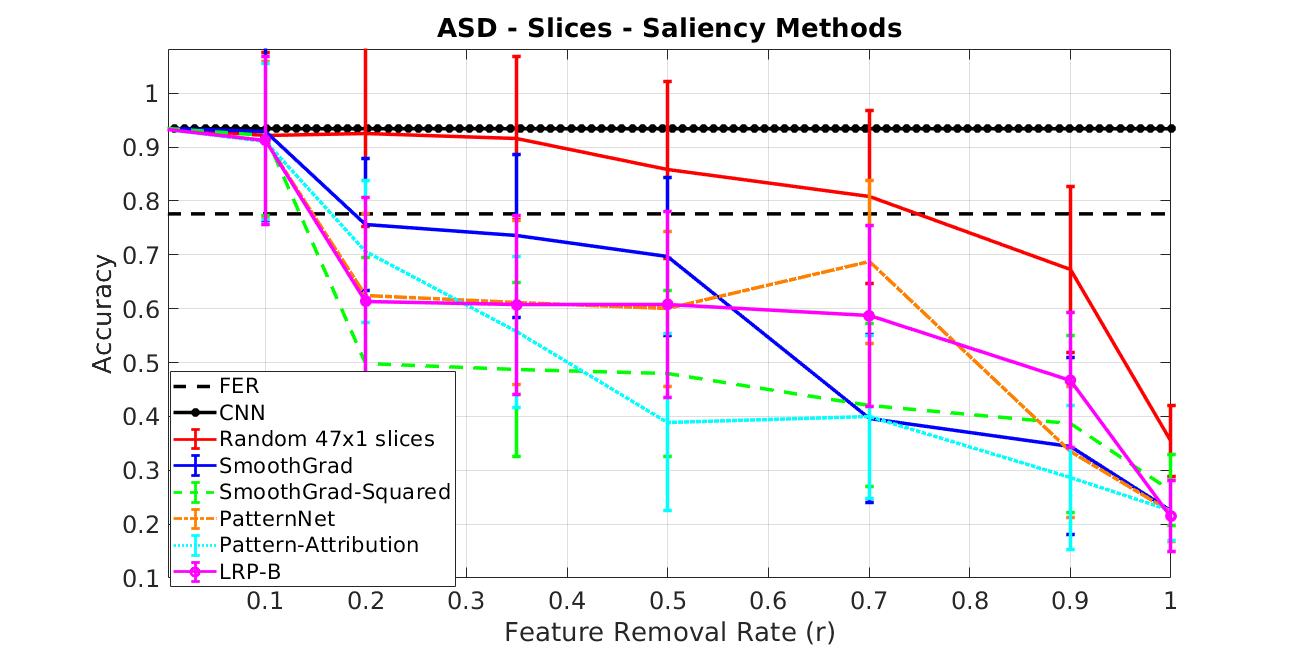}
   \label{fig_ASD_slices}
}
\caption{\scriptsize{Average accuracies and detriments comparison between all the saliency maps evaluated in this study. On Figures \ref{fig_TD_all} and \ref{fig_ASD_all}, and for the method-based slices on Figure Figures \ref{fig_TD_slices} and \ref{fig_ASD_slices}. TD plots are on the left and ASD on the right columns. Bars showed for each value of $r$ represent the standard-deviation of the set of accuracies wrapped for on each $r$ value and on each group - TD or ASD.}}
\label{accuracies_comparison}
\end{figure*}
Figures \ref{fig_TD_all} and \ref{fig_ASD_all} shows the accuracy detriment for all the XAI methods included in this study for TD and ASD respectively. Similarly, to illustrate the performance detriment obtained from the method-related 47$\times$1 baseline we show the accuracy detriment in Figure \ref{fig_TD_slices} and \ref{fig_ASD_slices} for TD and ASD.
Analyzing the differences between TD and ASD accuracies across the XAI methods we observed significant differences between ROAR-related accuracies and the random baselines. For Smooth-Grad  we found significant differences in comparison with the random baseline for $r$=0.7 (F(1,95)=3.31, p=0.00155) and r=0.9 (F(1,95)=2.65, p=0.0224) in TD.  For ASD we found differences in $r$=0.7 (F(1,79)=4.01, p=0.000331), and $r$=0.9 (F(179)=2.23, p=0.0338).  

Again for Smooth-Grad we found differences in comparison with the method-related slices in $r$=0.7 (F(1,95)=10.58,p=1.45e-6), and in $r$=0.9 (F(1,95)=7.33, p=0.00023) for TD. For ASD we found differences in $r$=0.7 (F(1,79)=11.11, p=2.28E-7), and in $r$=0.9 (F(1,79)=7.45, p=0.000148). For this particular method, all the method-related slice baseline accuracies were lower in comparison with the SmoothGrad-related performance detriment. Other differences in the accuracy detriment are more observable in low values of $r$. For instance, for Smooth-Grad Squared and $r$=0.2 (F(1,95)=10.99, p=0.0000274) for TD, and $r$=0.2 (F(1,79)=12.45, p=0.0000345) for ASD. 
The accuracy detriments associated with the Smooth-Grad Squared method-related slice baseline are lower than the XAI method itself. However, we did not find significant differences for TD (F(1,95)$\leq$2.58, p$>$0.1893) or ASD (F(1,79)$\leq$2.88, p$>$0.1910) for $r$ values different than 0.2.

For PatternNet we only found significant differences in comparison with the method-related slices baseline. For $r$=0.7 the XAI method always shows a lower accuracy for TD (F(1,95)=13.38, p=0.0001178), and for ASD (F(1,79)=20.45, p=2.77E-7). In $r$=0.9 we found differences for TD (F(1,95)=3.42, p=0.0267), and for ASD (F(1,79)=2.99, p=0.0321). For other values of $r$ we did not find differences after correction.
For Pattern-Attribution we found differences in r=0.2 for TD (F(1,95)=8.35, p=0.00327), and for ASD (F(1,79)=7.91, p=0.00899). We found other differences where the method-related slice baseline is showing a lower accuracy than the XAI method itself, specifically in $r$=0.5 for TD (F(1,95)=10.12, p=0.000224) and ASD (F(1,79)=9.88, p=0.003367). For other $r$ values we did not find any significant differences.

Comparing the ROAR accuracy detriments for the LRP-B preset method, we found significant differences in $r$=0.2 for TD (F(1,95)=3.56, p=0.00214), and for ASD (F(1,79)=3.66, p=0.00203). An important observation is that the accuracy detriment associated with the LRP-B method is only showing significant differences for ASD in $r$=0.7 (F(1,79)=3.91, p=0.000156). For other $r$ values we did not find any significant difference - (F(1,95)$\leq$1.99, p$>$0.2489).
\vspace{-0.2cm}
\section{Discussion}
DL pipelines are effective \cite{schirrmeister2017deep,torres2021facial,torresenhanced} at decoding facial emotions, but it is unclear what neural information is more important in this process. To address this, we compared XAI methods to determine their trustworthiness in this application \cite{samek2020toward}. We compared XAI saliency-maps (Figure \ref{fig:rel_maps}) generated from LRP-B, PatternNet, and Pattern-Attribution methods and identified features that are consistent with known patterns observed in EEG during facial emotion decoding for individuals with and without ASD \cite{webb2017face,webb2011developmental}.
Even though the previously-mentioned XAI methods identify features that are consistent with prior knowledge on EEG-based facial emotion decoding \cite{torres2021facial}, we observed some quantitative differences between the associated relevance patterns. Because of these differences, it is not clear which method best represents the neural-network facial emotion-decoding process. This motivates the usage of ROAR and random and method-slices baseline for quantifying the level of relevance of each of these XAI methods.ROAR yields a  reliable evaluation of what features of the EEG input are used to train the CNN for FER decoding. We observed that LRP, PatternNet and Pattern-Attribution identify the late time-ranges, after 500ms relative to the stimulus onset, as essential for correct facial emotion decoding, which is consistent with previous literature \cite{black2017mechanisms,dawson2005understanding,friedrich2015effective}. 

These analyses demonstrate that XAI methods can be applied to  recover features necessary for neural encoding.  However, this effect is only observed when less than 50\% of the relevant features are removed, as we observed in Figures 8.a and 8.c. While the best performing method differs based on the exact setting, we recommend using the SmoothGrad-Squared approach, since it is the easiest to implement, before using more complicated methods.

The results discussed above have shown that the patterns found by the XAI methods are meaningful. For PatternNet, Pattern-Attribution, Smooth-Grad Squared, and LRP-B when r=0.5 we see significant differences in time-ranges later than 500 ms. These differences are consistent with the late activation neural component in ASD when performing FER \cite{dawson2002neural}. The three time-ranges that were essential for facial emotion decoding in ASD were centered at 250, 600, and 1100 ms. These ranges were not essential for facial emotion decoding in TD. Instead, in TD subjects only two time ranges were essential for facial encoding - centered at 250, and 1200 ms. This suggests that EEG facial emotion decoding is done differently between groups, which is consistent with other EEG studies \cite{weitz2018towards,black2017mechanisms}.
The success of the CNN as a fully-personalized classifier -based on a LOTO cross-validation per subject- is an important novelty including XAI methods evaluation. This evaluation shows that it is possible to decode emotions reliably from the EEG of ASD populations when they perform FER. This confirms that there is intact    emotion information encoding in ASD from the EEG single trial/image level \cite{mayor2021interpretable}, understanding fully-personalized neural representations using XAI methods might be useful to develop more efficient data-driven ASD interventions \cite{fellous2019explainable}.
\vspace{-0.2cm}
\section{Conclusion}
The certainty analysis employed in this study suggests that the correctly decoded emotions/trials on our proposed EEG-based CNN pipeline are associated with relevance patterns that show high-relevance values on late time-ranges. In time-ranges between 500-1500ms after the stimulus onset we find significant differences associated with the relevance patterns and the ROAR binary masks obtained from some XAI methods,  such as, LRP-B, PatternNet and Pattern-Attribution. This effect is also observed when evaluating ROAR on those methods, thus suggesting that the EEG relevant features assessed with ROAR are particularly useful for obtaining a successful emotion recognition.

The differences in accuracy detriment observed after evaluating ROAR are important for supporting the differences observed between TD and ASD relevance maps. The more reliable XAI methods obtained after using ROAR and removing more than the 50\% of the relevant features are  precisely the LRP-B, PatternNet and Pattern-Attribution. These methods are the ones that show significant differences in the late-timing between TD and ASD. Specifically, those differences between TD and ASD observed in the more reliable XAI methods are consistent with the altered neural connectivity patterns observed when individuals with ASD process emotion from faces.

This study consolidates important findings in ASD and computational-neuroscience research. The ROAR evaluation can identify the more reliable and intuitively important features that can successfully decode an emotion from EEG activity. These distinguishable patterns are setting a more consistent and remarkable set of information that is precisely relevant for the emotion decoding. This represents a different and intact emotion information  encoding in individuals with ASD that can be efficiently extracted by the CNN. These results can re-define the current state-of-the-art of facial emotion-decoding pipelines and XAI. This supports, CNN as a perceptual-based classifier, which overcomes the behavioral/neural emotion comprehension deficits observed in individuals with ASD.

This study is the first, quantitatively speaking, to employ ROAR for evaluating robust XAI methods on EEG-based facial emotion recognition. This study is also the first to use EEG features for evaluating the reliability and correctness of current state-of-the-art XAI methods, including EEG trials from  ASD and non-ASD individuals.

%
\vspace{-0.22cm}
\section*{Acknowledgment}
\scriptsize{
We would like to express our thankful feelings and deep appreciation to Pieter-Jan Kindermans, PhD and Maximilian Alber, PhD, affiliated with Google Brain, for their extensive and receptive collaboration in the construction and evaluation of all the Explainable Artificial Intelligence (XAI) methods included in this study, and for their effort in the instruction and evaluation of the iNNvestigate package including EEG data \href{https://github.com/albermax/innvestigate}{https://github.com/albermax/innvestigate}. MDL was supported by the National Institute of Mental Health (Grant No.R01MH110585), grants from the Alan Alda Fund for Communication, the American Psychological Association, and Association for Psychological Science,  as  well  as  Fellowships  from  the  American  Psychological Foundation, Jefferson Scholars Foundation, and International Max Planck Research. We would like to thank Stony Brook Research Computing and Cyber-infrastructure, and the Institute for Advanced Computational Science at Stony Brook University for access to the SeaWulf computing system, which was made possible by a \$1.4 M National Science Foundation grant(\#1531492). Code for the CNN can be found in \href{https://github.com/meiyor/Deep-Learning-Emotion-Decoding-using-EEG-data-from-Autism-individuals}{https://github.com/meiyor/Deep-Learning-Emotion-Decoding-using-EEG-data-from-Autism-individuals}. Data for replication samples can be made available upon request to the corresponding authors. Tessa Clarkson was supported by the National Institute of Mental Health of the National Institutes of Health under Award Number [F31MH122091, 2020]. The content is solely the responsibility of the authors and does not necessarily represent the official views of the National Institutes of Health. Tessa Clarkson was also supported by the Temple University Dissertation Completion Grant, the Temple University Public Policy Lab Graduate Fellowship, American Psychological Association (APA) Dissertation Research Award, and the Dr. Phillip J Bersh Memorial Student Award.
}
\vspace{-0.1cm}
\bibliographystyle{IEEEtran}
\bibliography{references}

\begin{thebibliography}{10}
\providecommand{\url}[1]{#1}
\csname url@samestyle\endcsname
\providecommand{\newblock}{\relax}
\providecommand{\bibinfo}[2]{#2}
\providecommand{\BIBentrySTDinterwordspacing}{\spaceskip=0pt\relax}
\providecommand{\BIBentryALTinterwordstretchfactor}{4}
\providecommand{\BIBentryALTinterwordspacing}{\spaceskip=\fontdimen2\font plus
\BIBentryALTinterwordstretchfactor\fontdimen3\font minus
  \fontdimen4\font\relax}
\providecommand{\BIBforeignlanguage}[2]{{%
\expandafter\ifx\csname l@#1\endcsname\relax
\typeout{** WARNING: IEEEtran.bst: No hyphenation pattern has been}%
\typeout{** loaded for the language `#1'. Using the pattern for}%
\typeout{** the default language instead.}%
\else
\language=\csname l@#1\endcsname
\fi
#2}}
\providecommand{\BIBdecl}{\relax}
\BIBdecl

\bibitem{krizhevsky2012imagenet}
A.~Krizhevsky, I.~Sutskever, and G.~E. Hinton, ``Imagenet classification with
  deep convolutional neural networks,'' in \emph{Advances in neural information
  processing systems}, 2012, pp. 1097--1105.

\bibitem{sutskever2014sequence}
I.~Sutskever, O.~Vinyals, and Q.~V. Le, ``Sequence to sequence learning with
  neural networks,'' in \emph{Advances in neural information processing
  systems}, 2014, pp. 3104--3112.

\bibitem{he2017mask}
K.~He, G.~Gkioxari, P.~Doll{\'a}r, and R.~Girshick, ``Mask r-cnn,'' in
  \emph{Proceedings of the IEEE international conference on computer vision},
  2017, pp. 2961--2969.

\bibitem{prince2019evaluation}
J.~Prince, F.~Andreotti, and M.~De~Vos, ``Evaluation of source-wise missing
  data techniques for the prediction of parkinson’s disease using
  smartphones,'' in \emph{ICASSP 2019-2019 IEEE International Conference on
  Acoustics, Speech and Signal Processing (ICASSP)}.\hskip 1em plus 0.5em minus
  0.4em\relax IEEE, 2019, pp. 3927--3930.

\bibitem{mustafamultimodal}
H.~M. O'Leary, J.~M. Mayor, W.~E. Kaufmann, and M.~Sahin, ``Multimodal hand
  stereotypies detection in rett syndrome treatment using deep belief neural
  networks,'' \emph{2017 39th Annual International Conference of the IEEE
  Engineering in Medicine and Biology Society (EMBC)}, 2017.

\bibitem{andreotti2018multichannel}
F.~Andreotti, H.~Phan, N.~Cooray, C.~Lo, M.~T. Hu, and M.~De~Vos,
  ``Multichannel sleep stage classification and transfer learning using
  convolutional neural networks,'' in \emph{2018 40th Annual International
  Conference of the IEEE Engineering in Medicine and Biology Society
  (EMBC)}.\hskip 1em plus 0.5em minus 0.4em\relax IEEE, 2018, pp. 171--174.

\bibitem{liu2016emotion}
W.~Liu, W.-L. Zheng, and B.-L. Lu, ``Emotion recognition using multimodal deep
  learning,'' in \emph{International conference on neural information
  processing}.\hskip 1em plus 0.5em minus 0.4em\relax Springer, 2016, pp.
  521--529.

\bibitem{li2018cross}
H.~Li, Y.-M. Jin, W.-L. Zheng, and B.-L. Lu, ``Cross-subject emotion
  recognition using deep adaptation networks,'' in \emph{International
  Conference on Neural Information Processing}.\hskip 1em plus 0.5em minus
  0.4em\relax Springer, 2018, pp. 403--413.

\bibitem{weitz2018towards}
K.~Weitz, T.~Hassan, U.~Schmid, and J.~Garbas, ``Towards explaining deep
  learning networks to distinguish facial expressions of pain and emotions,''
  in \emph{Forum Bildverarbeitung 2018}.\hskip 1em plus 0.5em minus 0.4em\relax
  KIT Scientific Publishing, 2018, p. 197.

\bibitem{ghoshal2019estimating}
B.~Ghoshal, A.~Tucker, B.~Sanghera, and W.~L. Wong, ``Estimating uncertainty in
  deep learning for reporting confidence to clinicians when segmenting nuclei
  image data,'' in \emph{2019 IEEE 32nd International Symposium on
  Computer-Based Medical Systems (CBMS)}.\hskip 1em plus 0.5em minus
  0.4em\relax IEEE, 2019, pp. 318--324.

\bibitem{torres2021facial}
J.~M.~M. Torres, T.~Clarkson, K.~M. Hauschild, C.~C. Luhmann, M.~D. Lerner, and
  G.~Riccardi, ``Facial emotions are accurately encoded in the neural signal of
  those with autism spectrum disorder: A deep learning approach,''
  \emph{Biological Psychiatry: Cognitive Neuroscience and Neuroimaging}, pp.
  S2451--9022, 2021.

\bibitem{bosl2011eeg}
W.~Bosl, A.~Tierney, H.~Tager-Flusberg, and C.~Nelson, ``Eeg complexity as a
  biomarker for autism spectrum disorder risk,'' \emph{BMC medicine}, vol.~9,
  no.~1, p.~18, 2011.

\bibitem{castelhano2018stimulus}
J.~Castelhano, P.~Tavares, S.~Mouga, G.~Oliveira, and M.~Castelo-Branco,
  ``Stimulus dependent neural oscillatory patterns show reliable statistical
  identification of autism spectrum disorder in a face perceptual decision
  task,'' \emph{Clinical Neurophysiology}, vol. 129, no.~5, pp. 981--989, 2018.

\bibitem{jenke2014feature}
R.~Jenke, A.~Peer, and M.~Buss, ``Feature extraction and selection for emotion
  recognition from eeg,'' \emph{IEEE Transactions on Affective Computing},
  vol.~5, no.~3, pp. 327--339, 2014.

\bibitem{koelstra2011deap}
S.~Koelstra, C.~Muhl, M.~Soleymani, J.-S. Lee, A.~Yazdani, T.~Ebrahimi, T.~Pun,
  A.~Nijholt, and I.~Patras, ``Deap: A database for emotion analysis; using
  physiological signals,'' \emph{IEEE transactions on affective computing},
  vol.~3, no.~1, pp. 18--31, 2011.

\bibitem{fan2017eeg}
J.~Fan, E.~Bekele, Z.~Warren, and N.~Sarkar, ``Eeg analysis of facial affect
  recognition process of individuals with asd performance prediction leveraging
  social context,'' in \emph{2017 Seventh International Conference on Affective
  Computing and Intelligent Interaction Workshops and Demos (ACIIW)}.\hskip 1em
  plus 0.5em minus 0.4em\relax IEEE, 2017, pp. 38--43.

\bibitem{fan2017eeg2}
J.~Fan, J.~W. Wade, A.~P. Key, Z.~E. Warren, and N.~Sarkar, ``Eeg-based affect
  and workload recognition in a virtual driving environment for asd
  intervention,'' \emph{IEEE Transactions on Biomedical Engineering}, vol.~65,
  no.~1, pp. 43--51, 2017.

\bibitem{blankertz2011single}
B.~Blankertz, S.~Lemm, M.~Treder, S.~Haufe, and K.-R. M{\"u}ller,
  ``Single-trial analysis and classification of erp components—a tutorial,''
  \emph{NeuroImage}, vol.~56, no.~2, pp. 814--825, 2011.

\bibitem{winkler2011automatic}
I.~Winkler, S.~Haufe, and M.~Tangermann, ``Automatic classification of
  artifactual ica-components for artifact removal in eeg signals,''
  \emph{Behavioral and Brain Functions}, vol.~7, no.~1, p.~30, 2011.

\bibitem{o2017classification}
H.~M. O'Leary, J.~M. Mayor, C.-S. Poon, W.~E. Kaufmann, and M.~Sahin,
  ``Classification of respiratory disturbances in rett syndrome patients using
  restricted boltzmann machine,'' in \emph{2017 39th Annual International
  Conference of the IEEE Engineering in Medicine and Biology Society
  (EMBC)}.\hskip 1em plus 0.5em minus 0.4em\relax IEEE, 2017, pp. 442--445.

\bibitem{torres2013eeg}
J.~M.~M. Torres, ``Eeg signals classification using linear and non-linear
  discriminant methods,'' \emph{El Hombre y la M{\'a}quina}, no.~41, pp.
  71--80, 2013.

\bibitem{mayor2018t}
J.~M. Mayor~Torres, E.~Libsack, T.~Clarkson, C.~Keifer, G.~Riccardi, and
  M.~Lerner, ``Eeg-based single trial classification emotion recognition: A
  comparative analysis in individuals with and without autism spectrum
  disorder,'' \emph{International Society for Autism Research, INSAR 2018},
  vol.~85, no.~10, pp. S149--S150, 2018.

\bibitem{yang2015use}
H.~Yang, S.~Sakhavi, K.~K. Ang, and C.~Guan, ``On the use of convolutional
  neural networks and augmented csp features for multi-class motor imagery of
  eeg signals classification,'' in \emph{2015 37th Annual International
  Conference of the IEEE Engineering in Medicine and Biology Society
  (EMBC)}.\hskip 1em plus 0.5em minus 0.4em\relax IEEE, 2015, pp. 2620--2623.

\bibitem{ang2008filter}
K.~K. Ang, Z.~Y. Chin, H.~Zhang, and C.~Guan, ``Filter bank common spatial
  pattern (fbcsp) in brain-computer interface,'' in \emph{2008 IEEE
  International Joint Conference on Neural Networks (IEEE World Congress on
  Computational Intelligence)}.\hskip 1em plus 0.5em minus 0.4em\relax IEEE,
  2008, pp. 2390--2397.

\bibitem{lee2014classifying}
Y.-Y. Lee and S.~Hsieh, ``Classifying different emotional states by means of
  eeg-based functional connectivity patterns,'' \emph{PloS one}, vol.~9, no.~4,
  p. e95415, 2014.

\bibitem{blankertz2004bci}
B.~Blankertz, K.-R. Muller, G.~Curio, T.~M. Vaughan, G.~Schalk, J.~R. Wolpaw,
  A.~Schlogl, C.~Neuper, G.~Pfurtscheller, T.~Hinterberger \emph{et~al.}, ``The
  bci competition 2003: progress and perspectives in detection and
  discrimination of eeg single trials,'' \emph{IEEE transactions on biomedical
  engineering}, vol.~51, no.~6, pp. 1044--1051, 2004.

\bibitem{torresenhanced}
J.~M.~M. Torres, T.~Clarkson, E.~A. Stepanov, C.~C. Luhmann, M.~D. Lerner, and
  G.~Riccardi, ``Enhanced error decoding from error-related potentials using
  convolutional neural networks,'' \emph{40th International Engineering in
  Medicine and Biology Conference, EMBC 2018}, Jul. 2018.

\bibitem{schirrmeister2017deep}
R.~T. Schirrmeister, J.~T. Springenberg, L.~D.~J. Fiederer, M.~Glasstetter,
  K.~Eggensperger, M.~Tangermann, F.~Hutter, W.~Burgard, and T.~Ball, ``Deep
  learning with convolutional neural networks for eeg decoding and
  visualization,'' \emph{Human brain mapping}, vol.~38, no.~11, pp. 5391--5420,
  2017.

\bibitem{black2017mechanisms}
M.~H. Black, N.~T. Chen, K.~K. Iyer, O.~V. Lipp, S.~B{\"o}lte, M.~Falkmer,
  T.~Tan, and S.~Girdler, ``Mechanisms of facial emotion recognition in autism
  spectrum disorders: insights from eye tracking and electroencephalography,''
  \emph{Neuroscience \& Biobehavioral Reviews}, vol.~80, pp. 488--515, 2017.

\bibitem{kovalerchuk2018toward}
B.~Kovalerchuk and N.~Neuhaus, ``Toward efficient automation of interpretable
  machine learning,'' in \emph{2018 IEEE International Conference on Big Data
  (Big Data)}.\hskip 1em plus 0.5em minus 0.4em\relax IEEE, 2018, pp.
  4940--4947.

\bibitem{selvaraju2017grad}
R.~R. Selvaraju, M.~Cogswell, A.~Das, R.~Vedantam, D.~Parikh, and D.~Batra,
  ``Grad-cam: Visual explanations from deep networks via gradient-based
  localization,'' in \emph{Proceedings of the IEEE International Conference on
  Computer Vision}, 2017, pp. 618--626.

\bibitem{chattopadhay2018grad}
A.~Chattopadhay, A.~Sarkar, P.~Howlader, and V.~N. Balasubramanian,
  ``Grad-cam++: Generalized gradient-based visual explanations for deep
  convolutional networks,'' in \emph{2018 IEEE Winter Conference on
  Applications of Computer Vision (WACV)}.\hskip 1em plus 0.5em minus
  0.4em\relax IEEE, 2018, pp. 839--847.

\bibitem{kim2019saliency}
B.~Kim, J.~Seo, S.~Jeon, J.~Koo, J.~Choe, and T.~Jeon, ``Why are saliency maps
  noisy? cause of and solution to noisy saliency maps,'' \emph{arXiv preprint
  arXiv:1902.04893}, 2019.

\bibitem{smilkov2017smoothgrad}
D.~Smilkov, N.~Thorat, B.~Kim, F.~Vi{\'e}gas, and M.~Wattenberg, ``Smoothgrad:
  removing noise by adding noise,'' \emph{arXiv preprint arXiv:1706.03825},
  2017.

\bibitem{kindermans2017reliability}
P.-J. Kindermans, S.~Hooker, J.~Adebayo, M.~Alber, K.~T. Sch{\"u}tt,
  S.~D{\"a}hne, D.~Erhan, and B.~Kim, ``The (un) reliability of saliency
  methods,'' \emph{arXiv preprint arXiv:1711.00867}, 2017.

\bibitem{samek2020toward}
W.~Samek, G.~Montavon, S.~Lapuschkin, C.~J. Anders, and K.-R. M{\"u}ller,
  ``Toward interpretable machine learning: Transparent deep neural networks and
  beyond,'' \emph{arXiv preprint arXiv:2003.07631}, 2020.

\bibitem{binder2016layer}
A.~Binder, S.~Bach, G.~Montavon, K.-R. M{\"u}ller, and W.~Samek, ``Layer-wise
  relevance propagation for deep neural network architectures,'' in
  \emph{Information Science and Applications (ICISA) 2016}.\hskip 1em plus
  0.5em minus 0.4em\relax Springer, 2016, pp. 913--922.

\bibitem{montavon2018methods}
G.~Montavon, W.~Samek, and K.-R. M{\"u}ller, ``Methods for interpreting and
  understanding deep neural networks,'' \emph{Digital Signal Processing},
  vol.~73, pp. 1--15, 2018.

\bibitem{Kindermans2017}
P.-J. Kindermans, K.~T. Sch{\"u}tt, M.~Alber, K.-R. M{\"u}ller, D.~Erhan,
  B.~Kim, and S.~D{\"a}hne, ``Learning how to explain neural networks:
  Patternnet and patternattribution,'' \emph{arXiv preprint arXiv:1705.05598},
  2017.

\bibitem{adebayo2018sanity}
J.~Adebayo, J.~Gilmer, M.~Muelly, I.~Goodfellow, M.~Hardt, and B.~Kim, ``Sanity
  checks for saliency maps,'' in \emph{Advances in Neural Information
  Processing Systems}, 2018, pp. 9505--9515.

\bibitem{hooker2018evaluating}
S.~Hooker, D.~Erhan, P.-J. Kindermans, and B.~Kim, ``Evaluating feature
  importance estimates,'' \emph{arXiv preprint arXiv:1806.10758}, 2018.

\bibitem{dawson2005understanding}
G.~Dawson, S.~J. Webb, and J.~McPartland, ``Understanding the nature of face
  processing impairment in autism: insights from behavioral and
  electrophysiological studies,'' \emph{Developmental neuropsychology},
  vol.~27, no.~3, pp. 403--424, 2005.

\bibitem{dawson2007development}
G.~Dawson and R.~Bernier, ``Development of social brain circuitry in autism,''
  \emph{Human behavior, learning, and the developing brain: Atypical
  development}, pp. 28--56, 2007.

\bibitem{nowicki2000manual}
S.~Nowicki, ``Manual for the receptive tests of the diagnostic analysis of
  nonverbal accuracy 2,'' \emph{Atlanta, GA: Department of Psychology, Emory
  University}, 2000.

\bibitem{lord2000autism}
C.~Lord, E.~H. Cook, B.~L. Leventhal, and D.~G. Amaral, ``Autism spectrum
  disorders,'' \emph{Neuron}, vol.~28, no.~2, pp. 355--363, 2000.

\bibitem{dawson2002neural}
G.~Dawson, L.~Carver, A.~N. Meltzoff, H.~Panagiotides, J.~McPartland, and S.~J.
  Webb, ``Neural correlates of face and object recognition in young children
  with autism spectrum disorder, developmental delay, and typical
  development,'' \emph{Child development}, vol.~73, no.~3, pp. 700--717, 2002.

\bibitem{delorme2004eeglab}
A.~Delorme and S.~Makeig, ``Eeglab: an open source toolbox for analysis of
  single-trial eeg dynamics including independent component analysis,''
  \emph{Journal of neuroscience methods}, vol. 134, no.~1, pp. 9--21, 2004.

\bibitem{webb2015guidelines}
S.~J. Webb, R.~Bernier, H.~A. Henderson, M.~H. Johnson, E.~J. Jones, M.~D.
  Lerner, J.~C. McPartland, C.~A. Nelson, D.~C. Rojas, J.~Townsend
  \emph{et~al.}, ``Guidelines and best practices for electrophysiological data
  collection, analysis and reporting in autism,'' \emph{Journal of autism and
  developmental disorders}, vol.~45, no.~2, pp. 425--443, 2015.

\bibitem{bigdely2018finding}
N.~Bigdely-Shamlo, G.~Ibagon, C.~Kothe, and T.~Mullen, ``Finding the optimal
  cross-subject eeg data alignment method for analysis and bci,'' in \emph{2018
  IEEE International Conference on Systems, Man, and Cybernetics (SMC)}.\hskip
  1em plus 0.5em minus 0.4em\relax IEEE, 2018, pp. 1110--1115.

\bibitem{mognon2011adjust}
A.~Mognon, J.~Jovicich, L.~Bruzzone, and M.~Buiatti, ``Adjust: An automatic eeg
  artifact detector based on the joint use of spatial and temporal features,''
  \emph{Psychophysiology}, vol.~48, no.~2, pp. 229--240, 2011.

\bibitem{hyvarinen2000independent}
A.~Hyv{\"a}rinen and E.~Oja, ``Independent component analysis: algorithms and
  applications,'' \emph{Neural networks}, vol.~13, no. 4-5, pp. 411--430, 2000.

\bibitem{coates2012learning}
A.~Coates and A.~Y. Ng, ``Learning feature representations with k-means,'' in
  \emph{Neural networks: Tricks of the trade}.\hskip 1em plus 0.5em minus
  0.4em\relax Springer, 2012, pp. 561--580.

\bibitem{lee2018effectiveness}
C.~S. Lee, S.~H. Lam, S.~T. Tsang, C.~M. Yuen, and C.~K. Ng, ``The
  effectiveness of technology-based intervention in improving emotion
  recognition through facial expression in people with autism spectrum
  disorder: a systematic review,'' \emph{Review Journal of Autism and
  Developmental Disorders}, vol.~5, no.~2, pp. 91--104, 2018.

\bibitem{huang2018adversarially}
H.~Huang, D.~Li, Z.~Zhang, X.~Chen, and K.~Huang, ``Adversarially occluded
  samples for person re-identification,'' in \emph{Proceedings of the IEEE
  Conference on Computer Vision and Pattern Recognition}, 2018, pp. 5098--5107.

\bibitem{kingma2014adam}
D.~P. Kingma and J.~Ba, ``Adam: A method for stochastic optimization,''
  \emph{arXiv preprint arXiv:1412.6980}, 2014.

\bibitem{glorot2011deep}
X.~Glorot, A.~Bordes, and Y.~Bengio, ``Deep sparse rectifier neural networks,''
  in \emph{Proceedings of the fourteenth international conference on artificial
  intelligence and statistics}, 2011, pp. 315--323.

\bibitem{kindermans2017patternnet}
P.-J. Kindermans, K.~T. Sch{\"u}tt, M.~Alber, K.-R. M{\"u}ller, and
  S.~D{\"a}hne, ``Patternnet and patternlrp--improving the interpretability of
  neural networks,'' \emph{stat}, vol. 1050, p.~16, 2017.

\bibitem{alber2019innvestigate}
M.~Alber, S.~Lapuschkin, P.~Seegerer, M.~H{\"a}gele, K.~T. Sch{\"u}tt,
  G.~Montavon, W.~Samek, K.-R. M{\"u}ller, S.~D{\"a}hne, and P.-J. Kindermans,
  ``innvestigate neural networks!'' \emph{Journal of Machine Learning
  Research}, vol.~20, no.~93, pp. 1--8, 2019.

\bibitem{bach2015pixel}
S.~Bach, A.~Binder, G.~Montavon, F.~Klauschen, K.-R. M{\"u}ller, and W.~Samek,
  ``On pixel-wise explanations for non-linear classifier decisions by
  layer-wise relevance propagation,'' \emph{PloS one}, vol.~10, no.~7, p.
  e0130140, 2015.

\bibitem{szepessy1989existence}
A.~Szepessy, ``An existence result for scalar conservation laws using measure
  valued solutions.'' \emph{Communications in Partial Differential Equations},
  vol.~14, no.~10, pp. 1329--1350, 1989.

\bibitem{winkler2015influence}
I.~Winkler, S.~Debener, K.-R. M{\"u}ller, and M.~Tangermann, ``On the influence
  of high-pass filtering on ica-based artifact reduction in eeg-erp,'' in
  \emph{2015 37th Annual International Conference of the IEEE Engineering in
  Medicine and Biology Society (EMBC)}.\hskip 1em plus 0.5em minus 0.4em\relax
  IEEE, 2015, pp. 4101--4105.

\bibitem{webb2017face}
S.~J. Webb, E.~Neuhaus, and S.~Faja, ``Face perception and learning in autism
  spectrum disorders,'' \emph{The Quarterly Journal of Experimental
  Psychology}, vol.~70, no.~5, pp. 970--986, 2017.

\bibitem{cohen2017does}
M.~X. Cohen, ``Where does eeg come from and what does it mean?'' \emph{Trends
  in neurosciences}, vol.~40, no.~4, pp. 208--218, 2017.

\bibitem{powers2011evaluation}
D.~M. Powers, ``Evaluation: from precision, recall and f-measure to roc,
  informedness, markedness and correlation,'' 2011.

\bibitem{abdi2010holm}
H.~Abdi, ``Holm’s sequential bonferroni procedure,'' \emph{Encyclopedia of
  research design}, vol.~1, no.~8, pp. 1--8, 2010.

\bibitem{banerjee2018kolmogorov}
B.~Banerjee and B.~Pradhan, ``Kolmogorov--smirnov test for life test data with
  hybrid censoring,'' \emph{Communications in Statistics-Theory and Methods},
  vol.~47, no.~11, pp. 2590--2604, 2018.

\bibitem{webb2011developmental}
S.~J. Webb, E.~J. Jones, K.~Merkle, K.~Venema, J.~Greenson, M.~Murias, and
  G.~Dawson, ``Developmental change in the erp responses to familiar faces in
  toddlers with autism spectrum disorders versus typical development,''
  \emph{Child development}, vol.~82, no.~6, pp. 1868--1886, 2011.

\bibitem{friedrich2015effective}
E.~V. Friedrich, A.~Sivanathan, T.~Lim, N.~Suttie, S.~Louchart, S.~Pillen, and
  J.~A. Pineda, ``An effective neurofeedback intervention to improve social
  interactions in children with autism spectrum disorder,'' \emph{Journal of
  autism and developmental disorders}, vol.~45, no.~12, pp. 4084--4100, 2015.

\bibitem{mayor2021interpretable}
J.~M. Mayor-Torres, M.~Ravanelli, S.~E. Medina-DeVilliers, M.~D. Lerner, and
  G.~Riccardi, ``Interpretable sincnet-based deep learning for emotion
  recognition from eeg brain activity,'' \emph{arXiv preprint
  arXiv:2107.10790}, 2021.

\bibitem{fellous2019explainable}
J.-M. Fellous, G.~Sapiro, A.~Rossi, H.~Mayberg, and M.~Ferrante, ``Explainable
  artificial intelligence for neuroscience: Behavioral neurostimulation,''
  \emph{Frontiers in neuroscience}, vol.~13, p. 1346, 2019.

\end{thebibliography}

%
\begin{IEEEbiography}[{\includegraphics[width=1in,height=1.1in]{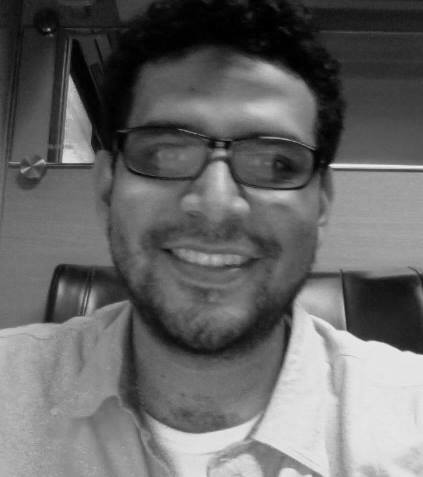}}]{Juan Manuel Mayor-Torres}
\scriptsize received the PhD in computer science in 2020 from the Department of Information Engineering and Compute (DISI), University of Trento, Italy.He received the BSc and MSc in Electrical Engineering from Pontificia Universidad Javeriana, Colombia in 2010 and 2014 respectively. He has worked as a Research Associate in the Department of Psychology of Cardiff University in 2020.He currently works as Postdoctoral Fellow in the School of Public Health, Physiotherapy \& Population Science of University College Dublin. He has been a reviewer in journals such as Expert Systems in Elsevier, MDPI Electronics and Applied Sciences, and in the International Conference of the IEEE Engineering in Medicine and Biology Society (EMBC) and ACM Interspeech. His research interests include Machine Learning, Emotion Recognition, Explainable AI, and Biomedical Signal Processing
\end{IEEEbiography}
\begin{IEEEbiography}[{\includegraphics[width=1in,height=1.1in]{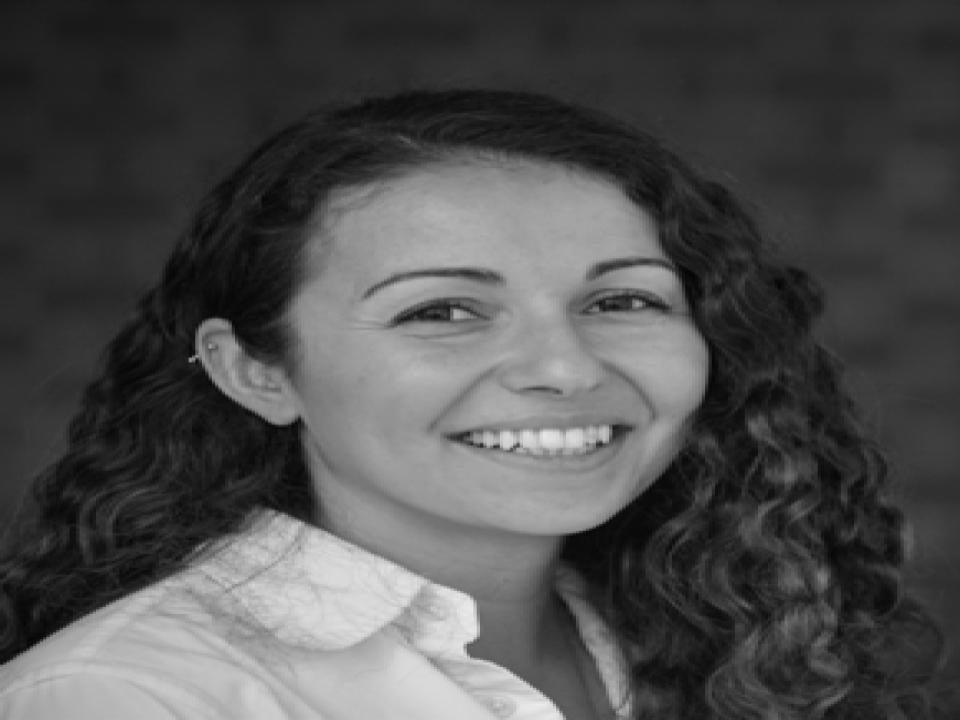}}]{Sara Medina-Devilliers}
\scriptsize received her PhD in Clinical Psychology at the University of Virginia in 2021 and her BA in Psychology and Neuroscience at Dartmouth College in 2012. Her research interests include investigating neural and physiological mechanisms underlying social relationships and social communication and their associations with health and well-being. She is currently a Postdoctoral Fellow at Boston Children’s Hospital
\end{IEEEbiography}
\begin{IEEEbiography}[{\includegraphics[width=1in,height=1.1in]{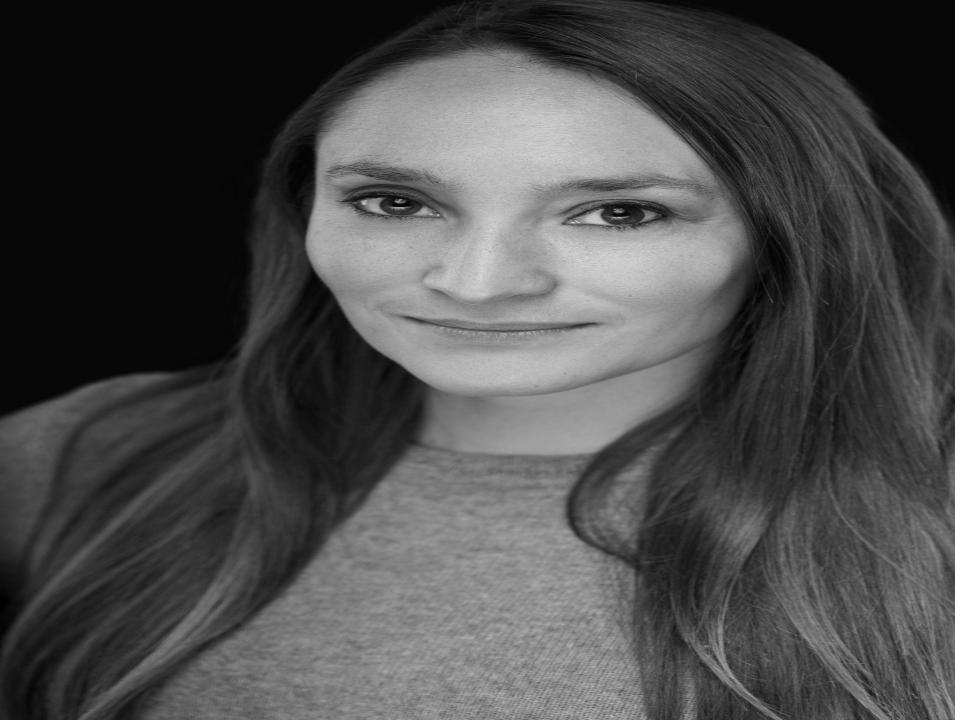}}]{Tessa Clarkson}
\scriptsize is a sixth-year doctoral candidate of the Clinical Psychology Ph.D. program at Temple University. She received her MA in Psychology at Stony Brook University in 2018 and her BS in Human Physiology at Boston University in 2011. She has worked as a reviewer in journals such as: the Journal of Autism and Developmental Disorders, Psychophysiology, Journal of Child Psychology and Psychiatry, Brain and Behavior, Plos One, Autism, and Social Cognitive and Affective Neuroscience. Her research interests include studying the underlying neural mechanisms of social learning, behavior, and cognition. She is also interested in how these mechanisms relate to the development and maintenance of psychopathology. 
\end{IEEEbiography}
\begin{IEEEbiography}[{\includegraphics[width=1.1in,height=1.1in]{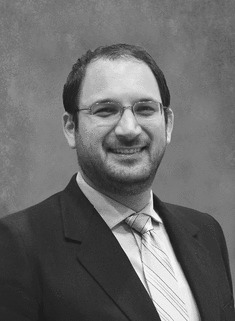}}]{Matthew D. Lerner}
\scriptsize is an Associate Professor of Psychology Psychiatry, \& Pediatrics in the Department of Psychology at Stony Brook University, where he directs the Social Competence and Treatment Lab. He is a Founder and Research Director of the Stony Brook Autism Initiative, and Co-Director of the Stony Brook LEND Center. He received his PhD in Clinical Psychology from the University of Virginia in 2013. Dr. Lerner's research focuses on understanding emergence and "real-world" implications of social problems in children and adolescents (especially those with Autism Spectrum Disorders [ASD]), as well as development, evaluation, and dissemination of novel, evidence-based approaches for ameliorating those problems. He has published more than 100 peer-reviewed articles and book chapters, he serves as Associate Editor of the Journal of Autism and Developmental Disorders, and on the Editorial Boards of 7 other academic outlets. He has presented at more than 150 national and international conferences on topics related to social development and developmental disorders. Dr. Lerner has received grants from organizations including the National Institutes of Health (NIH), the Health Resources \& Services Administration (HRSA), the Brain \& Behavior Research Foundation, the Simons Foundation, the American Psychological Association, and the American Academy of Arts \& Sciences. He has received several acknowledgments and awards, including the Biobehavioral Research Award for Innovative New Scientists (BRAINS) from the National Institute of Mental Health (NIMH), the Early Career Research Contributions Award from the Society for Research in Child Development (SRCD), the David Shakow Early Career Award for Distinguished Scientific Contributions to Clinical Psychology from the Society of Clinical Psychology (APA Division 12), the Sara S. Sparrow Early Career Research Award (APA Division 33), the Susan Nolen-Hoeksema Early Career Research Award from the Society for a Science of Clinical Psychology, the Richard “Dick” Abidin Early Career Award from the Society of Clinical Child and Adolescent Psychology (APA Division 53), Young Investigator Awards from the Brain \& Behavior Research Foundation (NARSAD) and the International Society for Autism Research, the Transformative Contributions Award from the Autism \& Developmental Disabilities SIG of the Association for Behavioral and Cognitive Therapies, and the Rising Star designation from the Association for Psychological Science. 
\end{IEEEbiography}
\begin{IEEEbiography}[{\includegraphics[width=1.1in,height=1.15in]{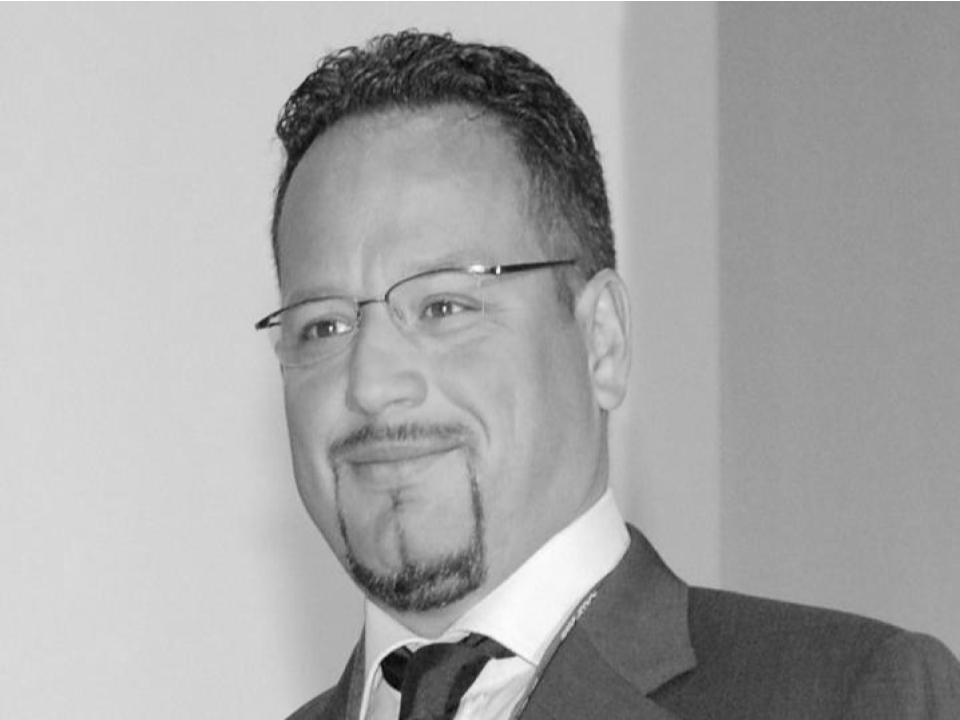}}]{Giuseppe Riccardi}
\scriptsize (M96-SM04-F10) is founder and director of the Signals and Interactive Systems Lab at University of Trento, Italy. He received his Laurea (MSEE) degree in Electrical Engineering and Master in Information Technology, from the University of Padua and Polytechnic of Milan (Italy), respectively. In 1995 he received his PhD in Electrical Engineering from the University of Padua, Italy. From 1993 to 2005, he was at AT\&T Bell Laboratories (USA) and then AT\&T Labs-Research (USA) where he worked in the Speech and Language Processing division. In 2005 joined the Department of Information Engineering and Computer Science (University of Trento). He has pioneered language modeling, understanding and human-machine dialogue system research. At AT\&T labs following the success in the 1994 DARPA spoken language understanding shared-task, he pioneered the well-known "How May I Help You?" research program which led to the first deployment of natural language dialogue systems. His team at University of Trento contributed to the IBM WATSON computer that won the Jeopardy! challenge in 2011. Prof. Riccardi has co-authored more than 230 papers and holds more than 90 patents. His research interests are language modeling, understanding, spoken/multimodal dialogue modeling and personal agents and affective computing. Prof. Riccardi has been on the scientific and organizing committee of IEEE, ISCA, ACMand ACL conferences. He has been a founder and Editorial Board member of the ACM Transactions of Speech and Language Processing ( now IEEE/ACM Transactions). He has been elected member of the IEEE SPS Speech Technical Committee (2005-2008). He is a member of ACL, ISCA, ACM and Fellow of IEEE and ISCA.
\end{IEEEbiography}
\end{document}